\documentclass[12pt]{article}

\usepackage{PRIMEarxiv}
\usepackage[a4paper,
            bindingoffset=0.2in,
            left=1in,
            right=1in,
            top=1in,
            bottom=1in,
            footskip=.25in]{geometry}
\usepackage{textcomp}
\usepackage{makecell}
\usepackage{longtable}
\usepackage[utf8]{inputenc} % allow utf-8 input
\usepackage[T1]{fontenc}    % use 8-bit T1 fonts
\usepackage[colorlinks=true,citecolor=blue]{hyperref}       % hyperlinks
\usepackage{url}            % simple URL typesetting
\usepackage{booktabs}       % professional-quality tables
\usepackage{amsfonts}       % blackboard math symbols
\usepackage{nicefrac}       % compact symbols for 1/2, etc.
\usepackage{microtype}      % microtypography
\usepackage{lipsum}
\usepackage{fancyhdr}       % header
\usepackage{graphicx}       % graphics
\usepackage[pagewise]{lineno}
\graphicspath{{media/}}     % organize your images and other figures under media/ folder

\usepackage{amsmath}
\newtheorem{remark}{Remark}
\numberwithin{equation}{section}
\usepackage{natbib}

\usepackage{enumitem}
\usepackage{setspace} 
\usepackage{multirow}
\usepackage[table,xcdraw]{xcolor}
\usepackage{float}
\usepackage[draft]{pgf}

\usepackage{stmaryrd}

\usepackage{multirow}

\renewcommand\arraystretch{1.2}
\usepackage[section]{placeins}
\usepackage{array}
\usepackage{booktabs}
\usepackage{subcaption}
\usepackage{etoolbox}
\AfterEndEnvironment{enumerate}{\vskip-\lastskip}
\usepackage[noend]{algpseudocode}
\usepackage[ruled]{algorithm}
\usepackage{chngcntr}
\usepackage[noabbrev,nameinlink]{cleveref}
\usepackage{rotating}
\usepackage{tikz}
\usepackage{amssymb}
 \DeclareMathOperator{\E}{\mathbb{E}}



\usepackage{titlesec}
\titlespacing*{\section}{0pt}{1ex}{1ex}
\titlespacing*{\subsection}{0pt}{1ex}{1ex}
\titlespacing*{\subsubsection}{0pt}{1ex}{1ex}
\usepackage[T1]{fontenc}
\usepackage{newtxmath,newtxtext}
\usepackage{array}  % For custom column types
\usepackage{graphicx}  % For \pbox if needed

\renewcommand{\arraystretch}{1.2} % Adjusts row spacing

\newcolumntype{L}[1]{>{\raggedright\arraybackslash}p{#1}}
\newcolumntype{C}[1]{>{\centering\arraybackslash}p{#1}}
\newcolumntype{R}[1]{>{\raggedleft\arraybackslash}p{#1}}

\allowdisplaybreaks

\title{\bf{Time-varying Parameter Tensor Vector Autoregression}}

\author{
Yiyong Luo %\\
   \And
  Jim E. Griffin%\\
 \AND{Department of Statistical Science, University College London, WC1E 6BT, UK}
}

\begin{document}
\maketitle

\begin{abstract}
Time-varying parameter vector autoregression provides a flexible framework to
capture structural changes within time series. However, when applied to high-
dimensional data, this model encounters challenges of over-parametrization and
computational burden. We address these challenges by building on recently proposed
Tensor VAR models to represent the time-varying coefficient matrix as a third-order
tensor with CANDECOMP/PARAFAC (CP) decomposition, yielding three model
configurations where different sets of components are specified as time-varying, each
offering distinct interpretations. To select the model configuration and the decomposition rank, we evaluate multiple variants of Deviance
Information Criterion (DIC) corresponding to the conditional and marginal DICs.
Our simulation demonstrates that a specific conditional DIC variant provides more
reliable results and accurately identifies true model configurations. We improve
the accuracy of rank selection by applying knee point detection to the DICs, rather
than defaulting to the minimum DIC value. Upon analyzing
functional magnetic resonance imaging data from story reading tasks, our selected
model configurations suggest time-varying dynamics while reducing the number
of parameters by over 90\% relative to standard VARs. Granger causality analysis
reveals directional brain connectivity patterns that align with narrative progression,
with various regions functioning as signal emitters or receivers at different time
points. 

\end{abstract}

% keywords can be removed
%\textbf{\normalisize{JEL Codes:}} C11, C13, C30, C53

%%%%%%%%%%%%% introduction %%%%%%%%%%%%%%
%%%%%%%%%%%%%%%%%%%%%%%%%%%%%%%%%%%%%%%%%
%%%%%%%%%%%%%%%%%%%%%%%%%%%%%%%%%%%%%%%%%
\setlength{\belowdisplayskip}{2pt} \setlength{\belowdisplayshortskip}{2pt}
\setlength{\abovedisplayskip}{2pt} \setlength{\abovedisplayshortskip}{2pt}
\newpage

\section{Introduction}

Time-varying parameter vector autoregression (TVP-VAR) is a multivariate time series model where coefficients evolve gradually according to random walks. Since its introduction in macroeconomics \citep{cogley2001evolving, cogley2005drifts, primiceri2005time}, TVP-VAR has become a methodological cornerstone in the field due to its ability to capture structural changes both flexibly and robustly \citep{nakajima2011time}. Beyond macroeconomics, TVP-VAR has been successfully applied in neuroscience \citep{havlicek2010dynamic} and finance \citep{geraci2018measuring}, where time series data exhibit non-stationarity. 

Applications of TVP-VARs to high-dimensional time series are challenging due to over-parameterization and computational burden. Consequently, empirical studies typically employ fewer than 10 variables to mitigate these issues. To retain the high-dimensional structure of the data, two strands have emerged in the literature. The first strand employs well-designed priors to maintain model parsimony. \cite{belmonte2014hierarchical}, \cite{pruser2021horseshoe}, 
\cite{eisenstat2016stochastic} and \cite{huber2021inducing} formulated TVP-VARs using the non-centered parameterization \citep{fruhwirth2010stochastic} and imposed shrinkage priors to both time-invariant parameters and the error terms governing time-varying dynamics. \cite{chan2023large} assigned a Bernoulli variable to each parameter to determine whether it should be modeled dynamically or remain static. The second strand implements dimension reduction techniques to decrease the number of unknown parameters. \cite{korobilis2013assessing} compressed high-dimensional data to low-dimensional factors and modeled these factors with a TVP-VAR. Similarly, \cite{chan2020reducing} and \cite{fischer2023general} reduced the parameter space by adopting factor-like structures to time-varying parameters. \cite{brune2022state} and \cite{cubadda2025time} proposed time-varying extensions of reduced-rank VAR \citep{carriero2011forecasting} and multivariate autoregressive index model \citep{reinsel1983some}, respectively. 

Our work falls within the second strand using dimension reduction, where we develop a time-varying parameter tensor VAR (TVP-TVAR). Firstly introduced by \cite{wang2022high}, tensor VAR (TVAR) expresses the time-invariant VAR coefficient matrix in terms of a third-order tensor and reduces the number of parameters via tensor decomposition. While \cite{luo2024bayesian} and \cite{chan2024large} have investigated Bayesian TVARs for modeling high-dimensional time series, their approaches focus on evolving variance-covariance matrices and static conditional mean. To leverage the advantage of TVARs within TVP-VARs, time-varying tensors would be necessary, yet research in this area remains scarce. Among the few existing studies, \cite{harris2021time} and \cite{chen2023discovering} constructed the tensors with time corresponding to one dimension without explicitly specifying the time variation; \cite{zhang2021bayesian} developed a model where a subset of elements within the tensor decomposition is active at each time point, controlled by a time-varying rank value - an unknown parameter in the decomposition. 

In this paper, we define a TVP-TVAR with the time-varying tensor and provide methods for posterior inference and model selection. In particular, we decompose the third-order tensor via the CANDECOMP/PARAFAC (CP) decomposition \citep{kiers2000towards} with a fixed rank, which represents the tensor as a sum of products of \textit{margins} (i.e. the elements in the tensor decomposition). The CP decomposition partitions margins into three \textit{loadings}: response, predictor and temporal. The response loading explains how data connects with a representation of past information, formulated by the predictor and temporal loadings. Each row of the predictor (temporal) loading determines how a specific variable (lag) contributes to this representation. Our model assumes one loading to be time-varying while maintaining the other two as time-invariant, resulting in three configurations. Although incorporating multiple time-varying loadings is possible, we restrict our model to these three configurations as they provide distinct interpretations of how temporal dynamics manifest in relationships between time series and their lags. Additionally, modeling only one time-varying loading is computationally efficient. By specifying the evolution of time-varying margins as random walks, we can treat the TVP-TVAR as a state-space model, facilitating straightforward Bayesian inference.

The TVP-TVAR model requires selecting an appropriate configuration (from one time-invariant and three time-varying options) and a CP decomposition rank. While existing tensor-structured models generally do not need to select model configuration, various methods have been applied in rank selection. \cite{guhaniyogi2021bayesian},  \cite{billio2023bayesian}, \cite{luo2024bayesian}, and \cite{fana2022bayesian}, among others, imposed shrinkage priors to eliminate redundant margins, thereby decreasing the rank from an initialized maximum. Alternatively, \cite{zhou2013tensor}, \cite{li2017parsimonious}, \cite{li2018tucker} and \cite{spencer2022parsimonious}, to name a few, employed information criteria, such as Bayesian information criterion \citep{schwarz1978estimating} and Deviance information criterion (DIC) \citep{spiegelhalter2002bayesian}, to determine the optimal rank. Given the prevalence of DIC in Bayesian analysis and its capacity to simultaneously select both configuration and rank with computational efficiency, we adopt DIC for model selection. 

Since the TVP-TVAR represents a state-space model with time-varying parameters being latent variables, several DIC variants established in latent variable models are available for model selection \citep{celeux2006deviance}. Choosing an appropriate DIC variant has received increasing attention in time series research \citep{chan2016observed, chan2018bayesian,li2020deviance} and other statistical domains \citep{ariyo2020bayesian,ariyo2022model,merkle2019bayesian}. We compare two types of DICs extensively investigated in the literature: conditional and marginal DICs. The conditional DIC evaluates the likelihood conditional on latent variables, while marginal DIC integrates them out. We specify two conditional and one marginal DICs based on the TVP-TVAR framework. While most references mentioned favor the marginal DIC due to concerns about uncertainty and complexity bias in conditional DICs, our simulation study reveals that one particular conditional DIC demonstrates advantages for TVP-TVARs — it yields more reliable results measured by lower Monte Carlo errors compared to alternative DICs, and accurately identifies the true model configuration. Based on these findings, we opt for this conditional DIC for model selection. Although higher ranks lead to lower chosen DICs, the improvement diminishes gradually as the rank increases, aligning with the finding in \cite{maity2021bayesian} that DIC distinguishes underfitted models but not overfitted ones. Thus, instead of determining the rank corresponding to the minimum DIC, we implement the "kneedle" algorithm \citep{satopaa2011finding} to a sequence of DICs across ranks for knee point detection, which improves the performance of identifying the optimal rank.

We demonstrate the utility of TVP-TVARs using functional magnetic resonance imaging (fMRI) data sets collected by \cite{wehbe2014simultaneously} from multiple subjects while reading a chapter in \textit{Harry Potter and the Sorcerer’s Stone} \citep{harry}. The result from model selection indicates that the dynamics of the fMRI data are time-varying, which reflects the conclusion of \cite{gaschler1997statistical} about the non-stationarity of the fMRI data. The CP decompositions with selected ranks manage to reduce the number of parameters by over 90\%, relative to standard VARs. The Granger causality analysis from the selected model shows that the number of directional brain connectivity is consistent with the narrative progression. Different regions of interest function as primary signal emitters or receivers at various time points.

The rest of the paper is organized as follows. Section \ref{Methodology} specifies the model framework. Section \ref{Posterior Computation} describes the posterior sampler of unknown parameters. Section \ref{Model and Rank Selection} discusses DIC variants and details our implementation of the "kneedle" algorithm. A Monte Carlo study in Section \ref{A Monte Carlo Study} supports one of the conditional DICs for model selection. Section \ref{Empirical Results} provides the empirical results of applying TVP-TVARs to the fMRI data.

%%%%%%%%%%%%% Tensor VAR %%%%%%%%%%%%%%
%%%%%%%%%%%%%%%%%%%%%%%%%%%%%%%%%%%%%%%%%
%%%%%%%%%%%%%%%%%%%%%%%%%%%%%%%%%%%%%%%%%
\section{Methodology} \label{Methodology}
\subsection{Tensor VAR}
A time-invariant tensor VAR with lag order $P$, denoted as TVAR($P$), uses a third-order tensor to store the coefficients in a VAR. In particular, the mathematical expression is written as 
\begin{align}
\boldsymbol{y}_t&=\boldsymbol{A}_1\boldsymbol{y}_{t-1}+...+\boldsymbol{A}_P\boldsymbol{y}_{t-P}+\boldsymbol{\epsilon}_t, \label{standard_VAR}\\
&= \boldsymbol{\mathcal{A}}_{(1)}\boldsymbol{x}_t+\boldsymbol{\epsilon}_t, \label{tensor_VAR}
\end{align}
where $\boldsymbol{y}_t\in\mathbb{R}^N$ denotes the multivariate data of interest at time point $t$, $\boldsymbol{x}_t=(\boldsymbol{y}^\prime_{t-1},...,\boldsymbol{y}^\prime_{t-P})^\prime\in \mathbb{R}^{NP}$ represents the $P$ lags of $\boldsymbol{y}_t$, $\boldsymbol{\mathcal{A}}\in\mathbb{R}^{N\times N\times P}$ is the third-order tensor, with $\boldsymbol{\mathcal{A}}_{i_1,i_2,p}$ corresponding to the ($i_1,\,i_2$) entry in $\boldsymbol{A}_p\in\mathbb{R}^{N\times N}$, for $i_1, \, i_2 = 1,\dots, N$ and $p=1,\dots, P$, $\boldsymbol{\mathcal{A}}_{(1)}=\left(\boldsymbol{A}_1,\dots, \boldsymbol{A}_P\right)\in \mathbb{R}^{N\times NP}$ is the mode-1 matricization of $\boldsymbol{\mathcal{A}}$, see Appendix \ref{appendix:Basic Notations and Definitions} for the definition. 

TVAR is especially useful in modeling high-dimensional time series because it can reduce the number of parameters via tensor decomposition. Among the tensor decompositions reviewed in \cite{kolda2009tensor}, we employ CANDECOMP/PARAFAC (CP) decomposition \citep{kiers2000towards} in this paper due to its simplicity. Specifically, the CP decomposition of $\boldsymbol{\mathcal{A}}$ is given by
\begin{equation}
\boldsymbol{\mathcal{A}}=\sum_{r=1}^R\boldsymbol{\mathcal{A}}^{(r)}=\sum_{r=1}^R\boldsymbol\beta^{(r)}_1\circ\boldsymbol\beta^{(r)}_3\circ\boldsymbol\beta^{(r)}_3,  \label{CP_decomposition}
\end{equation}
where $R$ is the decomposition rank, $\boldsymbol{\mathcal{A}}^{(r)}$ is a third-order tensor with dimensions identical to $\boldsymbol{\mathcal{A}}$, for $r=1,\dots,R$. The vectors $\boldsymbol{\beta}^{(r)}_1$, $\boldsymbol{\beta}^{(r)}_2\in\mathbb{R}^{N}$ and $\boldsymbol{\beta}^{(r)}_3\in\mathbb{R}^{P}$ are called \textit{margins}, which form $\boldsymbol{\mathcal{A}}^{(r)}$ through an outer product such that the $(i_1,i_2,i_3)$ entry in $\boldsymbol{\mathcal{A}}^{(r)}$ equals to $\boldsymbol\beta^{(r)}_{1,i_1}\boldsymbol\beta^{(r)}_{2,i_2}\boldsymbol\beta^{(r)}_{3,i_3}$ for $i_1,i_2=1,\dots,N$ and $i_3=1,\dots,P$ (see Appendix \ref{appendix:Basic Notations and Definitions} for the definition of outer product). We denote the CP decomposition in \labelcref{CP_decomposition} as $\boldsymbol{\mathcal{A}}=\llbracket \boldsymbol{B}_1,\boldsymbol{B}_2,\boldsymbol{B}_3\rrbracket_{\text{CP}}$, where $\boldsymbol{B}_j=\left(\boldsymbol\beta^{(1)}_j,\cdots,\boldsymbol\beta^{(R)}_j\right)\in\mathbb{R}^{I_j\times R}$ is referred to as \textit{loadings}, for $j=1,2,3$, $I_1=I_2=N$ and $I_3=P$. With an upper bound $R<N^2P/(2N\texttt{+}P)$, the number of parameters decreases from $N^2P$ in $\boldsymbol{\mathcal{A}}$ to $(2N\texttt{+}P)R$ in $\boldsymbol{B}_1$, $\boldsymbol{B}_2$ and $\boldsymbol{B}_3$. Thus, when considering the coefficient matrix alone, TVAR allows parameters to increase linearly with the number of variables rather than quadratically, as in standard VAR models. This low-rank structure in the CP decomposition introduces parsimony, alleviating the over-parameterization issue commonly encountered when analyzing high-dimensional time series.

The three loadings in the CP decomposition constitute the interpretation of a TVAR. Recall that \cite{wang2022high} and \cite{luo2024bayesian} referred to $\boldsymbol{B}_1$, $\boldsymbol{B}_2$ and $\boldsymbol{B}_3$ as response, predictor and temporal loadings, respectively. The terminology reflects their functional roles, and \cite{luo2024bayesian} provided a detailed discussion of these roles. To summarize: each row of the response loading explains how an individual variable in $\boldsymbol{y}_t$ connects to a representation of past information, $\boldsymbol{B^\prime}_2\boldsymbol{X}_t\boldsymbol{B}_3$, where $\boldsymbol{X}_t=(\boldsymbol{y}_{t-1},\dots,\boldsymbol{y}_{t-P})$; while each row of the predictor (temporal) loading determines how a specific variable (lag) contributes to this representation.

\subsection{Time-varying Parameter Tensor VAR}
\label{Time-varying Parameter Tensor VAR}
We incorporate temporal dynamics into  $\boldsymbol{\mathcal{A}}$ to formulate a time-varying parameter TVAR (TVP-TVAR)
\begin{equation}
\boldsymbol{y}_t=\boldsymbol{\mathcal{A}}_{t,(1)}\boldsymbol{x}_t+\boldsymbol{\epsilon}_t,\quad \boldsymbol{\epsilon}_t\sim\mathcal{N}\left(\mathbf{0}, \boldsymbol{\Omega}\right),\label{TVP_tensor_VAR}
\end{equation}
where $\boldsymbol{\mathcal{A}}_{t,(1)}=\boldsymbol{A}_t=(\boldsymbol{A}_{t,1},\dots,\boldsymbol{A}_{t,P})$ corresponds to the mode-1 matricization of the time-varying tensor $\boldsymbol{\mathcal{A}}_t$.  We define three CP decompositions of $\boldsymbol{\mathcal{A}}_t$ with the ranks being fixed over time: (1) $\boldsymbol{\mathcal{A}}_t=\llbracket \boldsymbol{B}_{t,1},\boldsymbol{B}_2,\boldsymbol{B}_3\rrbracket_{\text{CP}}$, (2) $\boldsymbol{\mathcal{A}}_t=\llbracket \boldsymbol{B}_{1},\boldsymbol{B}_{t,2},\boldsymbol{B}_3\rrbracket_{\text{CP}}$, (3) $\boldsymbol{\mathcal{A}}_t=\llbracket \boldsymbol{B}_{1},\boldsymbol{B}_{2},\boldsymbol{B}_{t,3}\rrbracket_{\text{CP}}$. Each decomposition sets one loading as time-varying while keeping the other two time-invariant. We refer to a TVP-TVAR with $P$ lags and the $j$-th loading being time-varying as TVP-TVAR$(P,j)$, for $j=1,2,3$.

While the CP decomposition can include more than one time-varying loading, we restrict it to the above three configurations mainly because each of them offers a distinct interpretation of how temporal dynamics manifest in the relationships between $\boldsymbol{y}_t$ and its lags. Extended from the interpretation of loadings in a TVAR, TVP-TVAR($P,1$) models a static construction of the past information, but the response of $\boldsymbol{y}_t$ to the past information evolves over time. Conversely, TVP-TVAR ($P,2$) and ($P,3$) operate under the opposite framework: the transformation of lagged values into the representation of past information varies temporally, while the relationship between $\boldsymbol{y}_t$ and this representation is fixed. In particular, these two decompositions allow time-varying effects from individual variables or lags to
the representation. Two additional considerations support our focus on TVP-TVAR($P,j$): identifiability constraints and computational efficiency. Regarding identifiability, Remark 1 establishes the identifiability conditions for the margins\footnote{While margins in TVP-TVAR($P,j$) are non-identifiable, the composite tensor formed by these margins maintains identifiability \citep{zhou2013tensor,guhaniyogi2017bayesian, zhang2021bayesian}.}. Notably, incorporating multiple time-varying loadings in the CP decomposition would cause both scaling and permutation transformations to be time-dependent, posing challenges in comparing model configurations. 

% which can be explained by rearranging $\labelcref{tensor_VAR}$ and $\labelcref{CP_decomposition}$ to
% \begin{equation}
% \boldsymbol{y}_t=\boldsymbol{B}_1\boldsymbol{\mathcal{I}}_{(1)}\text{vec}(\boldsymbol{B^\prime}_2\boldsymbol{X}_t\boldsymbol{B}_3)+\boldsymbol{\epsilon}_t,
% \label{tensorVAR_represent}
% \end{equation}
% where $\boldsymbol{\mathcal{I}}_{(1)}\in\mathbb{R}^{R\times R^2}$ is the mode-1 matricization of a third-order superdiagonal tensor $\boldsymbol{\mathcal{I}}$ with ones on non-zero entries (see the definition in Appendix ??), $\boldsymbol{X}_t=(\boldsymbol{y}_{t-1},\dots,\boldsymbol{y}_{t-P})$. The lagged values, $\boldsymbol{X}_t$, are transformed through the linear operation $\boldsymbol{B^\prime}_2\boldsymbol{X}_t\boldsymbol{B}_3$, which can be considered as a representation of past information before establishing its relationship with $\boldsymbol{y}_t$ via $\boldsymbol{B}_1$. As shown in \labelcref{tensorVAR_represent}, each row of the response loading explains how individual element in $\boldsymbol{y}_t$ connect to the past information, while each row of the predictor (temporal) loading determines how a specific predictor (lag) contributes to construct the past information. 
\begin{remark}
\textup{
    TVP-TVAR($P,j$) is identified up to scaling and permutation. Taking TVP-TVAR($P,1$) as an example. $\boldsymbol{\mathcal{A}}_t=\llbracket\boldsymbol{B}_{t,1},\, \boldsymbol{B}_2,\, \boldsymbol{B}_3\rrbracket_\text{CP}=\llbracket\boldsymbol{\tilde{B}}_{t,1},\, \boldsymbol{\tilde{B}}_2,\, \boldsymbol{\tilde{B}}_3\rrbracket_\text{CP}$, if loadings in the second CP decomposition satisfies:
\begin{enumerate}[noitemsep,topsep=0pt]
    \item Scaling: $\tilde{\boldsymbol{B}}_{t,1}=\boldsymbol{B}_{t,1}\boldsymbol{R}_1$, and $\tilde{\boldsymbol{B}}_{j^\prime}=\boldsymbol{B}_{j^\prime}\boldsymbol{R}_{j^\prime}$, for $j^\prime=2$ and 3. $\boldsymbol{R}_j$ is an $R$-by-$R$ diagonal matrix with $\prod_{j=1}^{3}\boldsymbol{R}_{j,(r,r)}=1$ for $r=1,\,\dots,\, R$, where $\boldsymbol{R}_{j,(r,r)}$ is the $r$-th diagonal term in $\boldsymbol{R}_{j}$.
    \item Permutation: $\tilde{\boldsymbol{B}}_{1,t}=\boldsymbol{B}_{1,t}\boldsymbol\Pi$ and $\tilde{\boldsymbol{B}}_{j^\prime}=\boldsymbol{B}_{j^\prime}\boldsymbol\Pi$, for $j=2$ and 3, and an arbitrary $R$-by-$R$ column-wise permutation matrix $\boldsymbol\Pi$.
\end{enumerate}
Margins in TVP-TVAR($P,2$) and TVP-TVAR($P,3$) are weakly identifiable under analogous conditions.
}
\end{remark}
\vspace{-0.3cm}

Throughout the rest of this paper, we denote the time-varying loading in TVP-TVAR ($P,j$) as $\boldsymbol{B}_{t,j}$ and the time-invariant ones as $\boldsymbol{B}_{j^\prime}$ for $j^\prime, j \in \{1,2,3\}$ and $j^\prime\neq j$. Following the standard structure in TVP-VAR \citep{primiceri2005time}, we model the evolution of $\boldsymbol{b}_{t,j}=\text{vec}\left(\boldsymbol{B}_{t,j}\right)$, for $t=2,\dots,T$, as a random walk
\begin{equation}
    \boldsymbol{b}_{t,j} = \boldsymbol{b}_{t-1,j} + \boldsymbol{\eta}_{t,j},  \quad \boldsymbol{\eta}_{t,j} \sim \mathcal{N}\left(\mathbf{0}, \boldsymbol{Q}_j\right), \label{TVP_margins}
\end{equation}
where $\boldsymbol{Q}_j=\text{diag}\left(q_{j,1}, \dots, q_{j,I_jR}\right)$ is a diagonal matrix, for $I_1, I_2=N$, $I_3=P$, $q_{j,k}$ follows an inverse-gamma prior, $ \mathcal{IG}\left(a_k, b_k\right)$\footnote{If a random variable $x$ follows $\mathcal{IG}\left(a,b\right)$, its probability density function is $\frac{b^a}{\Gamma(a)}(1/x)^{a+1}\exp (-b/x)$, where $\Gamma(\cdot)$ is the gamma function.}, for $k=1,\dots,I_jR$. We impose normal priors to $\boldsymbol{b}_{1,j}$ and $\boldsymbol{b}_{j^\prime}=\text{vec}(\boldsymbol{B}_{j^\prime})$, $\mathcal{N}\left(\mathbf{0},\boldsymbol{\Sigma}_{j}\right)$ and $\mathcal{N}\left(\mathbf{0},\boldsymbol{\Sigma}_{j^\prime}\right)$, respectively. Following the spirit of Minnesota-type prior \citep{litterman1979techniques}, which posits that the shorter lags contain more information than longer lags, we specify 
$\boldsymbol{\Sigma}_j=
    \begin{cases}
         \sigma^2\mathbf{I}_{NR}, &\,  j = 1 \text{ or } 2 \\
         \sigma^2\mathbf{I}_{R}\otimes \text{diag}\left(1, \frac{1}{2^2}, \dots, \frac{1}{P^2}\right),& \,\text{otherwise} 
         \end{cases}$, 
         where $\mathbf{I}_{NR}$ and $\mathbf{I}_R$ are identity matrices with dimensions specified in their subscripts\footnote{Note that $\Sigma_j$, for $j=1$ or 2, does not need to include estimated standard deviations like a Minnesota-type prior when the data is standardized.}. $\boldsymbol{\Sigma}_{j^\prime}$, the variance-covariance matrix of $\boldsymbol{b}_{j^\prime}$ share the same expression. $\boldsymbol{\Sigma}_3$ imposes an increasing shrinkage property to $\boldsymbol{B}_3$ (or the initialization of $\boldsymbol{B}_{t,3}$), wherein the shrinkage level increases with the loading row index, thereby prioritizing information from shorter lags over longer lags in the past information representation. These margin priors imply that the $(i_1,i_2,p)$ entry of the tensor initialization, decomposed by $\boldsymbol{B}_{1,j}$ and $\boldsymbol{B}_{j^\prime}$'s, has zero mean and variance $R\sigma^6/p^2$. Finally, $\boldsymbol{\Omega}$ follows an inverse-Wishart prior, $\boldsymbol{\Omega}\sim\mathcal{IW}\left(\nu, \boldsymbol{S}\right)$\footnote{If an $n$-by-$n$ positive definite matrix $\boldsymbol{Q}$ follows $\mathcal{IW}\left(\nu,\boldsymbol{S}\right)$, where $\nu$ is a scalar and $\boldsymbol{S}\in\mathbb{R}^{n\times n}$, then its probability density function is $\frac{\lvert S \rvert ^{\nu/2}}{2^{n \nu/2}\Gamma_n(\nu/2)}\lvert \boldsymbol{Q}\rvert ^{-\frac{\nu+n+1}{2}}e^{-\frac{1}{2}\text{tr}\left(\boldsymbol{S}\boldsymbol{Q}^{-1}\right)}$, where $\Gamma_n$ is the multivariate gamma function.}.

Two points are noteworthy if the time-varying margins follow random walks as defined in \labelcref{TVP_margins}. First, if $\sigma^2$ is a known parameter, margins are identifiable up to sign switching and permutation, instead of scaling and permutation described as in Remark 1. This is because if the two CP decompositions in Remark 1 provide the same tensor and satisfy the scaling condition, then $\tilde{\boldsymbol{b}}_{1,j}=\text{vec}\left(\tilde{\boldsymbol{B}}_{1,j}\right)$ follows a multivariate normal distribution with zero mean and variance-covariance matrix $\left(\boldsymbol{R}_j\otimes \boldsymbol{I}_{I_j}\right)\boldsymbol{\Sigma}_j\left(\boldsymbol{R}_j\otimes \boldsymbol{I}_{I_j}\right)^\prime$, which equals to $\boldsymbol{\Sigma}_j$ only when diagonal entries in $\boldsymbol{R}_j$ are 1 or -1. The same explanation also applies to margins in the time-invariant loadings, $\boldsymbol{B}_{j^\prime}$. According to \cite{luo2024bayesian}, the scaling indeterminacy of the CP decomposition is one of the sources of high autocorrelation in the MCMC samples of margins. To mitigate high autocorrelation from this source, we assume $\sigma^2$ to be known. Second, TVP-TVAR$(P,j)$ implies that the VAR coefficients evolve as random walks, which leads to a state-space representation
\begin{equation}
     \boldsymbol{y}_t=\left(\boldsymbol{I}_N\otimes \boldsymbol{x}^\prime_t\right)\boldsymbol{a}_t+\boldsymbol{\epsilon}_t, \quad \boldsymbol{a}_t=\boldsymbol{a}_{t-1}+\boldsymbol{\xi}_t, \label{equ:representation}
\end{equation}
where $\boldsymbol{a}_t=\text{vec}(\boldsymbol{A}^\prime_t)$. Suppose the model is TVP-TVAR$(P,1)$, then $\boldsymbol{\xi}_{t,i}$ (for $i=1,\dots,N^2P$) is independent to other elements in $\boldsymbol{\xi}_t$ and follow $\mathcal{N}\left(0,\sum_{r=1}^R\left(\boldsymbol{B}_{2,(i_2,r)}\right)^2\left(\boldsymbol{B}_{3,(i_3,r)}\right)^2q_{1,(r-1)+i_1}\right)$, for $i=NP(i_1-1)+N(i_3-1)+i_2$, where $\boldsymbol{B}_{j^\prime,(i_{j^\prime},r)}$ is the $(i_{j^\prime},r)$ entry of $\boldsymbol{B}_{j^\prime}$, for $i_1,i_2=1,\dots,N$ and $i_3=1,\dots,P$. If the model is TVP-TVAR$(P,2)$ or TVP-TVAR$(P,3)$, one can change the three terms in the normal distribution accordingly.

%%%%%%%%%%%%%% Bayesian inference %%%%%%%%%%%%%%
%%%%%%%%%%%%%%%%%%%%%%%%%%%%%%%%%%%%%%%%%
%%%%%%%%%%%%%%%%%%%%%%%%%%%%%%%%%%%%%%%%%

\section{Posterior Computation} \label{Posterior Computation}

We now turn to the MCMC algorithm to estimate unknown parameters in \labelcref{TVP_tensor_VAR} and \labelcref{TVP_margins} given the priors specified in the previous section and a fixed rank $R$. We adopt a Gibbs sampler to sample $\{\boldsymbol{b}_{1:T,j}, \boldsymbol{b}_{j^\prime}, \boldsymbol{Q}_j, \boldsymbol{\Omega}\}$. The sampler cycles through the following steps.

%Note that $\sigma^2$ in $\boldsymbol{\Sigma_j}$, for $j=1,\dots, 3$, is set as a known parameter to circumvent the rank indeterminacy that would arise in our model selection (see the next section) when the rank is unspecified\footnote{We demonstrate the rank indeterminacy as follows: the prior variance of the $(i_1,i_2,p)$ entry of the tensor initialization, $R\sigma^6/p^2$, equals to $\tilde{R}\tilde{\sigma}^6/p^2$ if $R=a\tilde{R}$ and $\sigma^6=\frac{1}{a}\tilde{\sigma}^6$, for any non-zero scalar $a$ such that $R$ and $\tilde{R}$ are integers. The same indeterminacy exists in the prior variance of coefficients in the TVAR.  Consequently, when employing evaluation metrics for rank selection, distinct ranks may yield identical metric values, hindering the determination of optimal rank choice.}. 

\textbf{Step 1.} We sample $\boldsymbol{b}_{1:T,j}$ using the forward filtering backward sampling algorithm \citep{fruhwirth1994data, carter1994gibbs}. This approach is feasible because, analogous to the rearrangements in TVARs, see \cite{wang2022high}, TVP-TVAR$(P,j)$ can be rearranged as 
\begin{equation}
\boldsymbol{y}_t=\boldsymbol{Z}_{t,j}\boldsymbol{b}^*_{t,j}+\epsilon_t
\label{equ:TVAR_represent}
\end{equation}
where $\boldsymbol{Z}_{t,1}=\left(\boldsymbol{x}^\prime_t\left(\boldsymbol{B}_3\otimes \boldsymbol{B}_2\right)\boldsymbol{\mathcal{I}}^\prime_{(1)}\right)\otimes \boldsymbol{I}_N$, $\boldsymbol{Z}_{t,2}=\boldsymbol{B}_1\boldsymbol{\mathcal{I}}_{(1)}\left((\boldsymbol{B}^\prime_3\boldsymbol{X}^\prime_t)\otimes \boldsymbol{I}_R\right)$, $\boldsymbol{Z}_{t,3}=\boldsymbol{B}_1\boldsymbol{\mathcal{I}}_{(1)}$ $\left(\boldsymbol{I}_R\otimes(\boldsymbol{B}^\prime_2\boldsymbol{X}_t)\right)$, $\boldsymbol{\mathcal{I}}\in\mathbb{R}^{R\times R \times R}$ is a superdiagonal tensor with ones on non-zero entries (the definition of superdiagonal tensor is in Appendix \ref{appendix:Basic Notations and Definitions}), $\boldsymbol{b}^*_{t,j}={\boldsymbol{b}}_{t,j}$ for $j=1$ and 3, $\boldsymbol{b}^*_{t,2}=\text{vec}\left(\boldsymbol{B}^\prime_{t,2}\right)$. Incorporating this equation with the random walk of $\boldsymbol{b}_{t,j}$ yields a linear state-space model. 

\textbf{Step 2.}  $\boldsymbol{b}_{j^\prime}$ comprises two blocks corresponding to loading specified in the previous section, with each block having similar full conditionals. Based on \labelcref{equ:TVAR_represent}, the full conditionals of $\boldsymbol{b}_{j^\prime}$, for $j^\prime=1$ or 3, or that of $\boldsymbol{b}^*_{j^\prime}$, for $j=2$, is $\mathcal{N}\left(\bar{\boldsymbol{\mu}}_{j^\prime},\bar{\boldsymbol{\Sigma}}_{j^\prime}\right)$ with
\begin{equation*}
\overline{\boldsymbol\Sigma}^{-1}_{j^\prime}= {\boldsymbol\Sigma}^{-1}_{j^\prime}+\sum_{t=1}^T\boldsymbol{Z}^\prime_{t,j^\prime}\boldsymbol{\Omega}^{-1}\boldsymbol{Z}_{t,j^\prime},\quad \bar{\boldsymbol{\mu}}_{j^\prime}= \overline{\boldsymbol\Sigma}_{j^\prime}\sum_{t=1}^T\boldsymbol{Z}^\prime_{t,j^\prime}\boldsymbol{\Omega}^{-1}{\boldsymbol{y}}_t,
\end{equation*}
where $\boldsymbol{B}_{j}$ in $\boldsymbol{Z}_{t,j^\prime}$ changes to $\boldsymbol{B}_{t,j}$.

\textbf{Step 3.} The by-product of samples from Step 1 and 2 is the time-varying tensor $\boldsymbol{\mathcal{A}}_t$. The variance-covariance matrix $\boldsymbol{\Omega}$ is sampled given $\boldsymbol{\mathcal{A}}_{t,(1)}$ and $\boldsymbol{y}_{1:T}$ from 
\begin{equation*}
    \mathcal{IW}\left(T+\nu, \sum_{t=1}^T\left(\boldsymbol{y}_t-\boldsymbol{\mathcal{A}}_{t,(1)}\boldsymbol{x}_t\right) \left(\boldsymbol{y}_t-\boldsymbol{\mathcal{A}}_{t,(1)}\boldsymbol{x}_t\right)^\prime+\boldsymbol{S}\right).
\end{equation*}

\textbf{Step 4.} Sample $q_{j,k}$, for $k=1,\dots, I_jR$, from 
\begin{equation*}
    \mathcal{IG}\left(a_k+\frac{T}{2}, b_k+\frac{1}{2}\sum_{t=2}^T\left(\boldsymbol{b}_{t,j,k}-\boldsymbol{b}_{t-1, j,k}\right)^2\right),
\end{equation*}
where $\boldsymbol{b}_{t,j,k}$ is the $k$-th element in $\boldsymbol{b}_{t,j}$.

%%%%%%%%%%%%% Model selection %%%%%%%%%%%%%%
%%%%%%%%%%%%%%%%%%%%%%%%%%%%%%%%%%%%%%%
%%%%%%%%%%%%%%%%%%%%%%%%%%%%%%%%%%%%%%%
\section{Model Selection}
\label{Model and Rank Selection}

Given the TVP-TVAR framework and a data set, we need to select a model configuration to describe the time variation of loadings and a rank value. There are 4 configurations corresponding to TVAR$(P)$ and TVP-TVAR$(P,j)$, for $j=1,2,3$, and we select the rank from 1 to $R^*$, a pre-defined upper bound. While existing tensor-structured models generally do not select model configuration, rank selection has been extensively investigated. \cite{guhaniyogi2017bayesian} and \cite{luo2024bayesian} imposed shrinkage priors to margins with a rank of $R^*$ and shrank the rank by removing redundant margins from the model. This approach requires only one MCMC simulation, rather than multiple simulations with different rank values, thus enhancing computational efficiency; however, given the complexity of the TVP-TVAR, we opt not to impose any additional shrinkage prior. Alternatively, one can select an evaluation metric and incrementally increase the rank from 1 until the metric no longer indicates improvement in model performance. One advantage of this approach compared to imposing shrinkage priors mentioned is that we can select both model configuration and rank simultaneously. Several evaluation metrics are available for model selection, including Bayes factor (BF) \citep{jeffreys1935some} and Deviance information criterion (DIC) \citep{spiegelhalter2002bayesian}. The DIC offers practical advantages, as it can be more readily evaluated from MCMC outputs compared to the BF\footnote{Beyond the scope of this paper, the DIC has additional advantages over BF. Notably, the latter suffers Jeffreys–Lindley paradox (see detailed description in \cite{robert2014jeffreys}) and is not well-defined with improper priors, i.e. the corresponding marginal likelihood is arbitrary up to a multiplicative constant, whereas the former does not have these issues.}. Given the widespread adoption of DIC in both Bayesian time series \citep{chan2016observed,chan2018bayesian,bai2015identification,li2020deviance} and tensor-structured models \citep{guhaniyogi2021bayesian,spencer2022parsimonious}, we use this metric for model selection.

To facilitate further discussion about DIC, we briefly introduce it in the TVP-TVAR framework. The deviance of parameter $\boldsymbol{\theta}$, $D(\boldsymbol{\theta})$, is defined as $D(\boldsymbol{\theta}) = -2\log p\left(\boldsymbol{y}_{1:T}\mid \boldsymbol{\theta}\right)+2\log h\left(\boldsymbol{y}_{1:T}\right)$, where $h\left(\boldsymbol{y}_{1:T}\right)$ is a fully-specified standardizing term which only depends on $\boldsymbol{y}_{1:T}$. The DIC balances goodness-of-fit and model complexity by a summation of two terms
\begin{equation*}
\text{DIC}\left(\boldsymbol{\theta}\right)=\overline{D(\boldsymbol{\theta})}+p_D,
\end{equation*}
where $\overline{D(\boldsymbol{\theta})}=-2\E_{\boldsymbol{\theta}}\left(\log p\left(\boldsymbol{y}_{1:T}\mid \boldsymbol{\theta}\right)\right)+2\log h\left(\boldsymbol{y}_{1:T}\right)$ is the posterior mean deviance and $p_D$ denotes the effective number of parameters. In particular, $p_D$ is the difference between posterior mean deviance and deviance of posterior mean, $\overline{D(\boldsymbol{\theta})}-D(\hat{\boldsymbol{\theta}})$. Since $h\left(\boldsymbol{y}_{1:T}\right)$ is independent of $\boldsymbol{\theta}$, one can set $h\left(\boldsymbol{y}_{1:T}\right)$ as 1 and write the DIC as
\begin{equation}
\text{DIC}\left(\boldsymbol{\theta}\right)=-4\E_{\boldsymbol{\theta}}\left(\log p\left(\boldsymbol{y}_{1:T}\mid\boldsymbol{\theta}\right)\right)+2\log p\left(\boldsymbol{y}_{1:T}\mid\hat{{\boldsymbol{\theta}}}\right), \label{conditional_DIC}
\end{equation}
where we approximate the first term by the MCMC samples of $\boldsymbol{\theta}$ and plug the sample mean $\hat{\boldsymbol{\theta}}$ into the second term.

Since the TVP-TVAR can be written as state-space models, as described in \labelcref{equ:representation} and \labelcref{equ:TVAR_represent}, $\boldsymbol{\boldsymbol{\mathcal{A}}}_{1:T}$ and $\boldsymbol{b}_{t,j}$ serve as latent variables in these two respective equations. Following \cite{celeux2006deviance}, if $\boldsymbol{\theta}$ includes latent variables, we regard the DIC in \labelcref{conditional_DIC} as conditional DIC because the likelihood is conditional on these latent variables. This conditional DIC has two variants based on the specification of $\boldsymbol{\theta}$. The first variant, denoted as $\text{DIC}^{c,1}$, is formulated with $\boldsymbol{\theta}=\{\boldsymbol{\mathcal{A}}_{1:T}, \boldsymbol{\Omega}\}$. The second variant, $\text{DIC}^{c,2}$, corresponds to the specification where $\boldsymbol{\theta}=\{\boldsymbol{b}_{1:T,j}, \boldsymbol{b}_{j^\prime}, \boldsymbol{\Omega}\}$. These two variants are distinct due to the difference in their deviances of posterior mean, $p\left(\boldsymbol{y}_{1:T}\mid\hat{{\boldsymbol{\theta}}}\right)$. Specifically, both conditional DIC variants need to compute the time-varying tensor based on $\hat{{\boldsymbol{\theta}}}$. Take the TVP-TVAR$(P,1)$ as an example, these tensors correspond to $\mathbb{E}\left[\llbracket\boldsymbol{B}_{t,1},\, \boldsymbol{B}_2,\, \boldsymbol{B}_3\rrbracket_\text{CP}\right]$ and $\llbracket\mathbb{E}\left[\boldsymbol{B}_{t,1}\right],\, \mathbb{E}\left[\boldsymbol{B}_2\right],\, \mathbb{E}\left[\boldsymbol{B}_3\right]\rrbracket_\text{CP}$ in $\text{DIC}^{c,1}$ and $\text{DIC}^{c,2}$, respectively, and these expressions can differ because the loadings may have posterior correlation. An alternative DIC widely studied in latent variable models is marginal DIC, denoted as $\text{DIC}^m$, which takes the form of \labelcref{conditional_DIC} with $\boldsymbol{\theta}=\{\boldsymbol{b}_{j^\prime}, \boldsymbol{Q}_j, \boldsymbol{\Omega}\}$. This DIC marginalizes $\boldsymbol{\mathcal{A}}_{1:T}$ and $\boldsymbol{b}_{1:T,j}$ in in $\text{DIC}^{c,1}$ and $\text{DIC}^{c,2}$, respectively, and uses an integrated likelihood
$p\left(\boldsymbol{y}_{1:T}\mid\boldsymbol{\theta}\right)=\int p\left(\boldsymbol{y}_{1:T}\mid \boldsymbol{\mathcal{A}}_{1:T},\boldsymbol{\Omega}\right)p\left(\boldsymbol{\mathcal{A}}_{1:T}\mid\boldsymbol{b}_{j^\prime},\boldsymbol{Q}_j\right)d\boldsymbol{\mathcal{A}}_{1:T}=\int p\left(\boldsymbol{y}_{1:T}\mid \boldsymbol{b}_{1:T,j},\boldsymbol{b}_{j^\prime},\boldsymbol{\Omega}\right)p\left(\boldsymbol{b}_{1:T,j}\mid 
\boldsymbol{b}_{j^\prime},\boldsymbol{Q}_j\right)d\boldsymbol{{b}}_{1:T,j}$ to evaluate model performance. A closed form expression for $\log p\left(\boldsymbol{y}_{1:T}\mid\boldsymbol{\theta}\right)$ is available in Appendix \ref{appendix:Marginal DIC}. 

Various studies favor marginal DIC over conditional ones for two reasons. Firstly, conditional DICs potentially exhibits more uncertainty than marginal DIC since it depends on the latent variables \citep{chan2016fast, merkle2019bayesian}. Secondly, the conditional DICs tend to choose the most complex model even if the data generating process is less complex \citep{millar2009comparison, chan2016fast}. However, we show that the concern about the first limitation can be invalid without a specific condition, and the second limitation of conditional DICs can be addressed by knee point detection.

For the first limitation about the uncertainty of DICs, we emphasize that this uncertainty stems from the posterior uncertainty of $\boldsymbol{\theta}$. Due to the indeterminacy of the CP decomposition, as discussed in Remark 1, margins defined in Section \ref{Time-varying Parameter Tensor VAR} are identifiable up to sign switching and permutation, resulting in highly autocorrelated margin Markov chains\footnote{Although the autocorrelation arising from the scaling indeterminacy has been mitigated, autocorrelation due to sign switching and permutation still occurs.}. The consequence of this autocorrelation is that margins present high sample variance within a single MCMC simulation, and their sample means exhibit high variation given multiple simulations (see Figure \ref{fig:hist_se}, \ref{fig:traceplots} and \ref{fig:boxplot} in the next section for an illustration). $\text{DIC}^{c,1}$ and $\text{DIC}^{m}$, which incorporate margins, inherently demonstrate greater uncertainty compared to $\text{DIC}^{c,1}$, whose parameters of interest do not suffer from this indeterminacy issue. Therefore, we prefer $\text{DIC}^{c,1}$ for model selection.

\begin{figure}[!htbp]
     \centering
     \begin{subfigure}[b]{0.35\textwidth}
         \centering
\includegraphics[width=\textwidth]{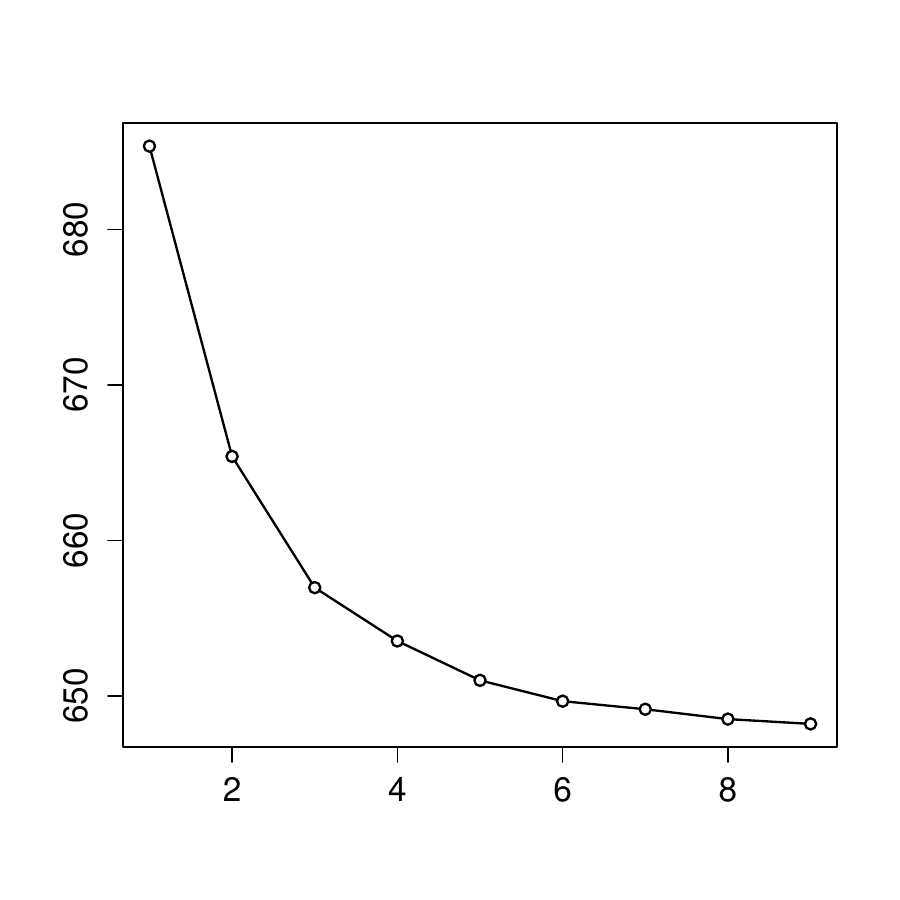}
         \caption{Unnormalized.}
         \label{fig:conditional_DIC_curve}
     \end{subfigure}
     \hspace{0.3cm}
     \begin{subfigure}[b]{0.35\textwidth}
         \centering
         \includegraphics[width=\textwidth]{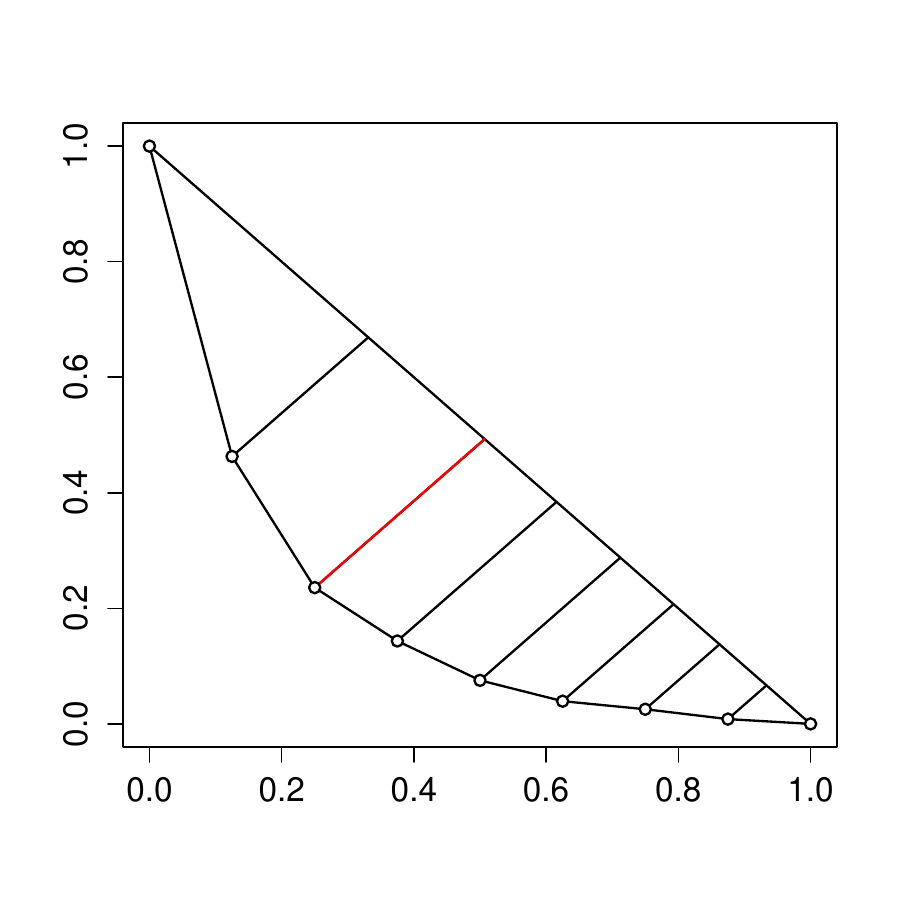}
         \caption{Normalized.}
         \label{fig:conditional_DIC_curve_normalized}
     \end{subfigure}
        \caption{Example of unnormalized (left) and normalized (right) conditional DICs (presented in circles) computed from the data set generated from TVP-TVAR($3,1$) with a rank of 3. The straight lines connecting the circles and $y=-x+1$ represent the distance between them, with the red line representing the maximum distance.}
        \label{fig:conditional_DIC_curves}
\end{figure}

We regard the model complexity about the second limitation in two facets: model configuration and rank choice. For the former, TVP-TVAR($P,1$) and TVP-TVAR($P,2$) are more complex due to more parameters included relative to those in TVP-TVAR($P,3$) and time-invariant TVAR. According to the Monte Carlo study about $\text{DIC}^{c,1}$ in Section \ref{A Monte Carlo Study}, this conditional DIC does not favor a more complex model configuration over the one that generates the data. Next, we move to the rank choice. While higher ranks do yield lower $\text{DIC}^{c,1}$, the model improvement gradually diminishes when the rank exceeds a certain value, based on the simulation results in the next section (see Figure \ref{fig:conditional_DIC_curve}). This reflects the finding in \cite{maity2021bayesian} that the DIC tends to distinguish underfitted models but not overfitted models\footnote{The DIC described in \cite{maity2021bayesian} corresponds to \labelcref{conditional_DIC}, but it cannot be considered as the conditional DIC because their model does not have any latent structure.}. Therefore, a useful tool to prevent $\text{DIC}^{c,1}$ from choosing the highest available rank is knee point detection, which detects the point of maximum curvature in a function.

 Multiple definitions of maximum curvature result in different algorithms for knee point detection. For example, \cite{zhao2008knee} defined an angle-based curvature to select the number of clusters using a Bayesian information criterion curve; \cite{satopaa2011finding} proposed "kneedle" to find the maximum distance between a normalized curve and a straight line (the detailed definition will be provided). We choose "kneedle" for knee point detection because it has a clearer visualization of maximum curvature and gives higher accuracy in detecting knees compared to the angle-based one. Specifically, locating the maximum curvature in a discrete sequence $\text{DIC}^{c,1}_\mathcal{M}=\left(\text{DIC}^{c,1}_{\mathcal{M},1},\dots, \text{DIC}^{c,1}_{\mathcal{M},R^*}\right)$, which stores the $\text{DIC}^{c,1}$'s with model configuration $\mathcal{M}$ and different rank values, involves two steps. The first step is to normalize each conditional DIC to $\widetilde{\text{DIC}}^{c,1}_{\mathcal{M},R}=\left(\text{DIC}^{c,1}_{\mathcal{M},R}-\text{DIC}^{c,1}_{\mathcal{M},R_\text{min}}\right)/\left(\text{DIC}^{c,1}_{\mathcal{M},R_\text{max}}-\text{DIC}^{c,1}_{\mathcal{M},R_\text{min}}\right)$, where $R_\text{min}$ and $R_\text{max}$ represent the ranks associated with the minimum and maximum DICs, and convert the corresponding rank $R$ to $(R-1)/(R^*-1)$. The curve in Figure \ref{fig:conditional_DIC_curve_normalized} represents these normalized DICs computed using the simulation data in Section \ref{A Monte Carlo Study}. Then, we identify the point with the maximum curvature as that with the maximum distance to the line which passes points $(0,\widetilde{\text{DIC}}^{c,1}_{\mathcal{M},1})$ and $(1,\widetilde{\text{DIC}}^{c,1}_{\mathcal{M},R^*})$. As illustrated in Figure \ref{fig:conditional_DIC_curve_normalized}, this maximum curvature occurs at the third point, so we select the rank as 3, which is the true rank value defined in the next section.

%%%%%%%%%%%%% Simulation %%%%%%%%%%%%%%
%%%%%%%%%%%%%%%%%%%%%%%%%%%%%%%%%%%%%%%
%%%%%%%%%%%%%%%%%%%%%%%%%%%%%%%%%%%%%%%

\section{A Monte Carlo Study} \label{A Monte Carlo Study}
%%%%%%%%%%%%% data and implementation %%%%%%%%%%%%%%
\subsection{Data and Implementation}

This section demonstrates the utility of $\text{DIC}^{c,1}$ (henceforth referred to as DIC for brevity) in model selection. We generate 100 3-variable datasets from each of four model configurations: TVAR(3), TVP-TVAR($3,j$), for $j=1,2,3$. Each data set contains 200 observations and the corresponding tensors are generated from CP decompositions with a rank of 3. We set $\sigma^2$, the multiplier in the margin variance, as 0.5, resulting in prior variances of 0.375, 0.094, and 0.042 for the coefficients in $\boldsymbol{A}_1$, $\boldsymbol{A}_2$. $\boldsymbol{A}_3$ (or their corresponding initializations in time-varying models), respectively. $\boldsymbol{\Omega}$ is sampled from the inverse-Wishart distribution stated in Section \ref{Time-varying Parameter Tensor VAR}, with $\nu=6$ and $\boldsymbol{S}=\mathbf{I}_3$, so that the prior mean of $\boldsymbol{\Omega}$ is $0.5\mathbf{I}_3$. $\boldsymbol{Q}_j$, the variance-covariance matrix in the random walk process of time-varying margins, is a diagonal matrix with all non-zero element being 0.01.

With $N=P=3$, the maximum rank that can be selected is 9, as $R\leq\{N^2, NP\}$ \citep{kolda2009tensor}. Therefore, there are 36 (9\texttimes 4) possible models to select from for each data set. We estimate parameters in TVP-TVAR models using the algorithm described in Section \ref{Posterior Computation}, while for TVAR models, we modify Step 2 in the algorithm to incorporate three blocks of loadings rather than two, followed by implementation of the remaining two steps in the algorithm. To compute the DIC of each model and data set, we run 10 parallel Markov chains, each having 10,000 iterations with 1,000 burn-in. Following parameter estimation, we calculate the DIC for knee point detection by averaging the DIC from each of the 10 runs. The 
"kneedle" algorithm then helps determine the rank choice for each of the four model configurations. After recording the four corresponding DICs, we select the model configuration with the lowest value.

%%%%%%%%%%%%% Rank Selection %%%%%%%%%%%%%%
\subsection{Simulation Result} \label{Simulation Result}

\begin{figure}[!htbp]
     \centering
     \begin{subfigure}[b]{0.3\textwidth}
         \centering
         \includegraphics[width=\textwidth]{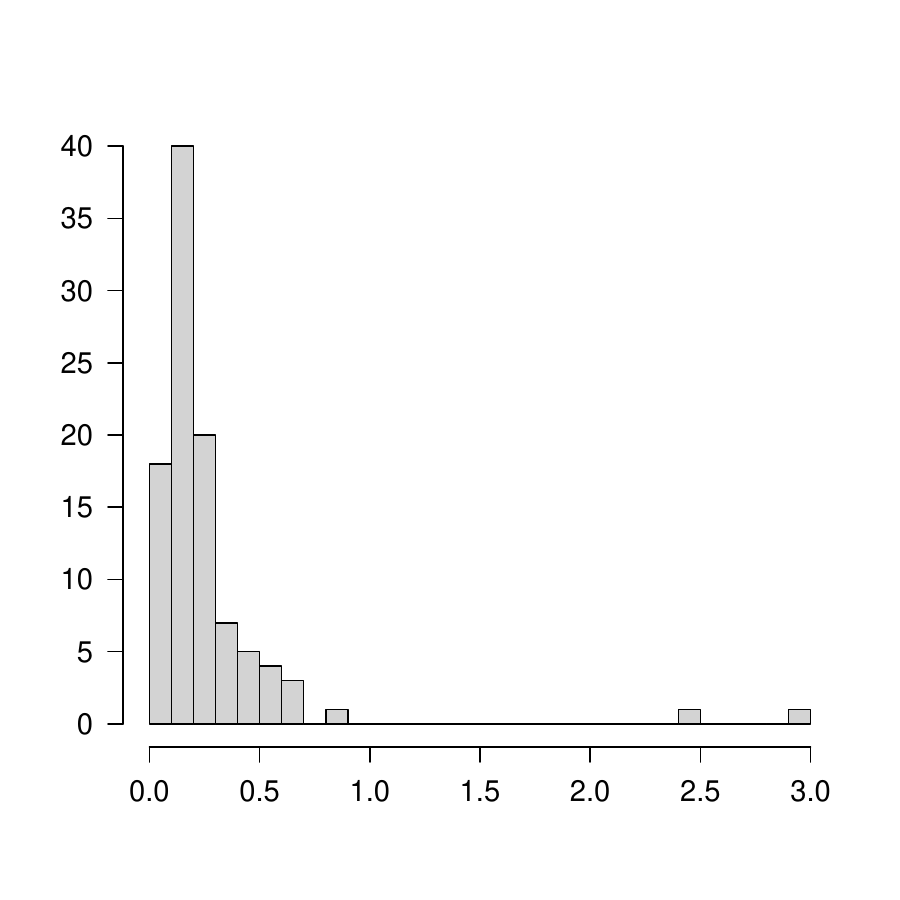}
         \caption{$\text{DIC}^{c,1}$.}
         \label{fig:se_conditional}
     \end{subfigure}
     \begin{subfigure}[b]{0.3\textwidth}
         \centering
\includegraphics[width=\textwidth]{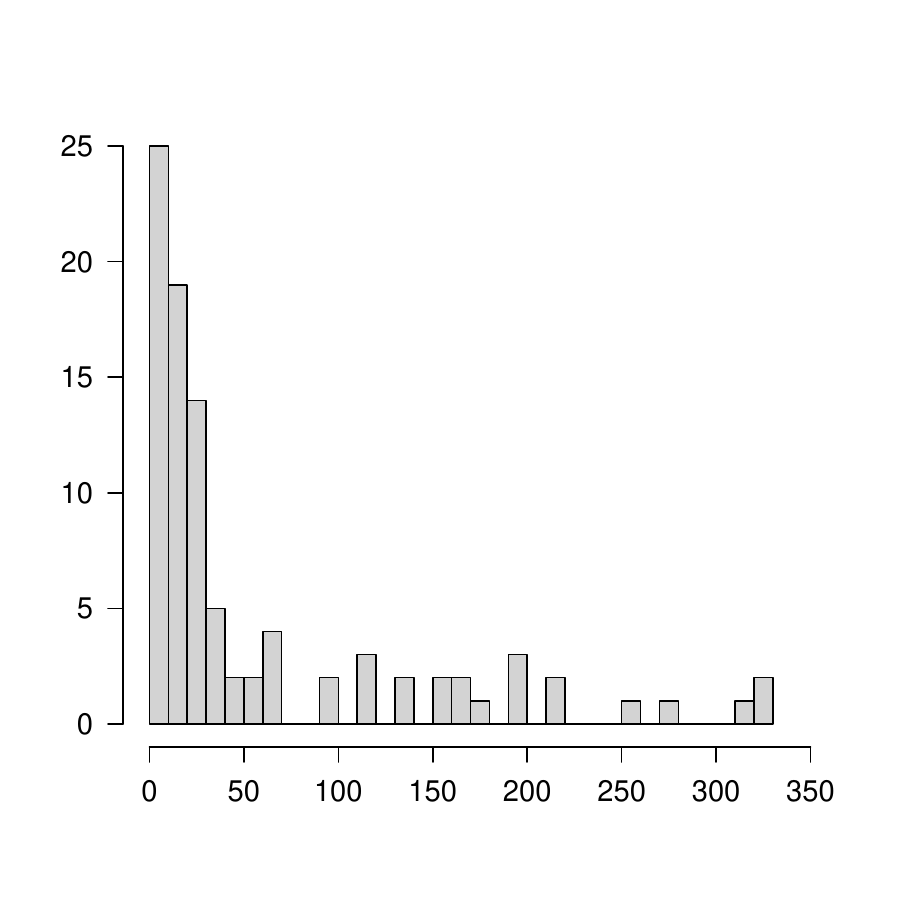}
         \caption{$\text{DIC}^{c,2}$.}
         \label{fig:se_conditional_1}
     \end{subfigure}
     \begin{subfigure}[b]{0.3\textwidth}
         \centering
         \includegraphics[width=\textwidth]{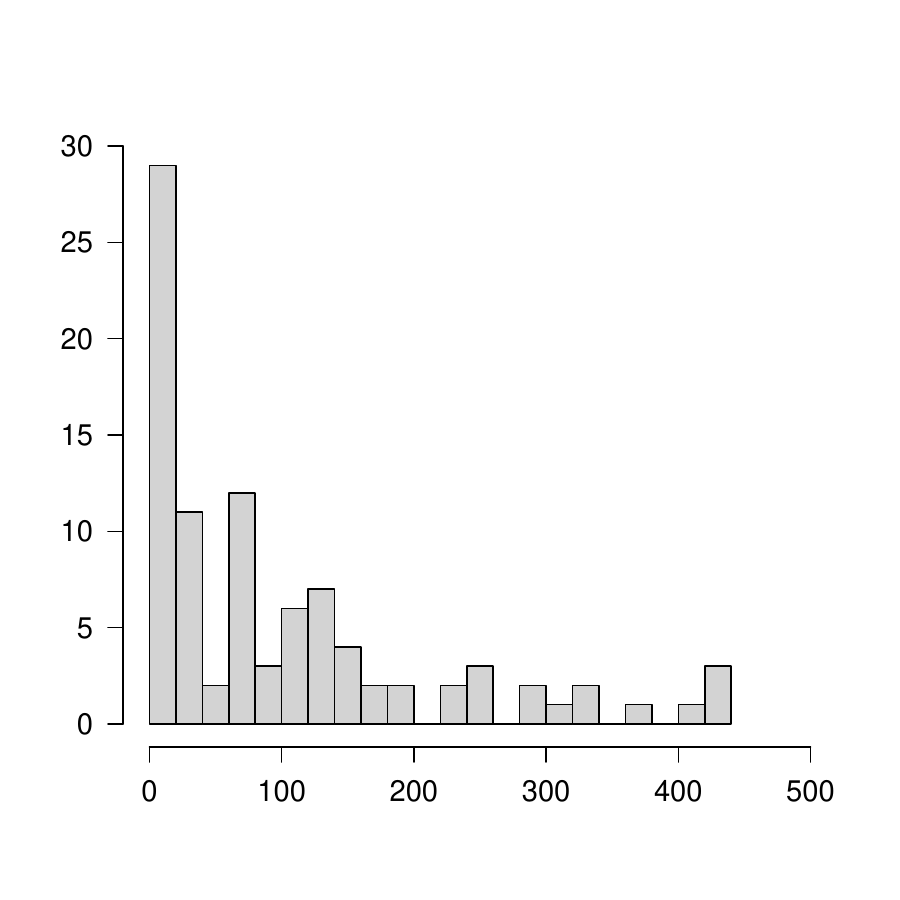}
         \caption{$\text{DIC}^{m}$}
         \label{fig:se_marginal}
     \end{subfigure}
     \hfill
        \caption{Histograms of Monte Carlo errors of conditional and marginal DICs, computed using data sets generated from TVP-TVAR($3,1$). The model configuration and rank associated with these DICs align with the true data generating process. The middle and right panels restrict the display of Monte Carlo errors to a maximum of 350 and 500, respectively, which accounts for 93 data sets. Each Monte Carlo error is scaled by $1/\sqrt{10}$ for visualization clarity.}
        \label{fig:hist_se}
\end{figure}

First, we compare the Monte Carlo errors corresponding to all the DIC variants mentioned in Section \ref{Model and Rank Selection}, and show the reason why we prefer $\text{DIC}^{c,1}$. The Monte Carlo error of a data set is calculated as the sample standard deviation of the 10 DICs obtained from the 10 parallel MCMC runs. A lower Monte Carlo error indicates that the corresponding DIC variant is more reliable because it has lower uncertainty. As illustrated in Figure \ref{fig:se_conditional}, most Monte Carlo errors of $\text{DIC}^{c,1}$ are below 1, whereas those in Figure \ref{fig:se_conditional_1} and \ref{fig:se_marginal} range from 0 to 500, with at most 10 being smaller than 1. These findings support our decision to use $\text{DIC}^{c,1}$ for model selection in the TVP-TVAR framework. 

\begin{figure}[htp]
    \centering
    
    % First row: two subfigures
    \begin{minipage}{0.48\textwidth}
        \centering
        \includegraphics[width=\textwidth]{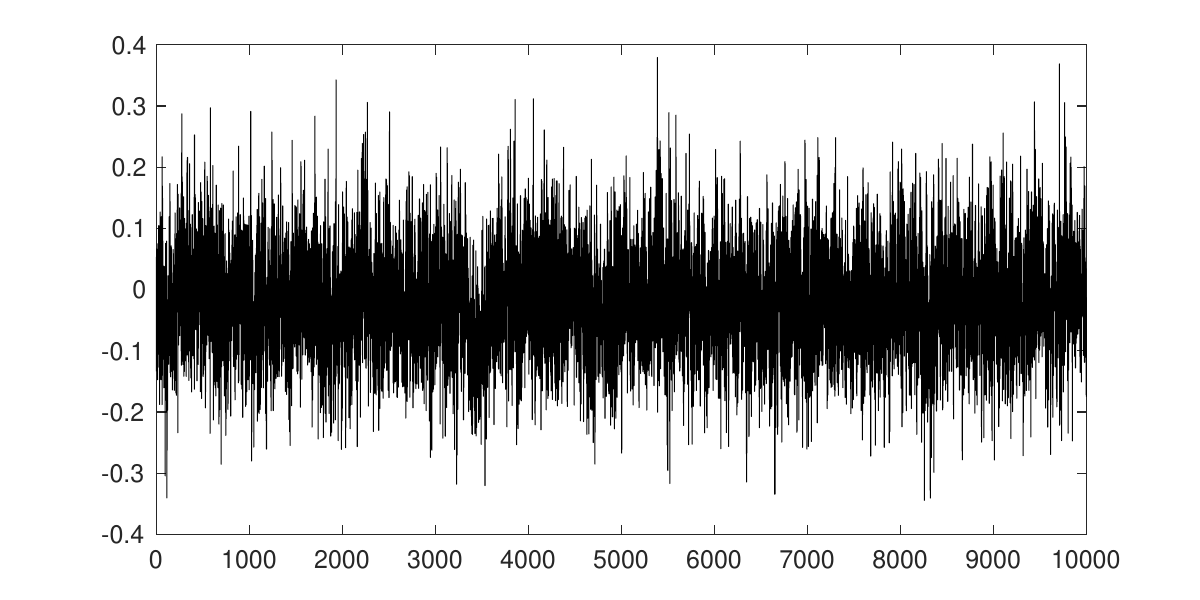} % Add your figure path here
        \subcaption{$\boldsymbol{A}_{t,(1,1)}$.}
    \end{minipage}%
    \hfill
    \begin{minipage}{0.48\textwidth}
        \centering
        \includegraphics[width=\textwidth]{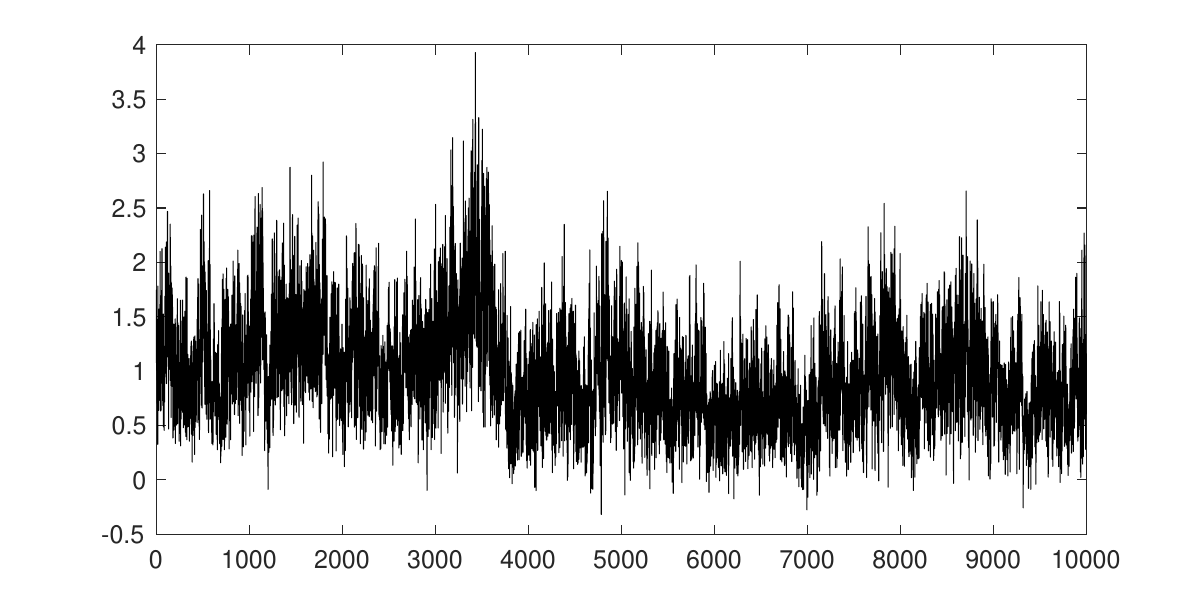} % Add your figure path here
        \subcaption{$\boldsymbol{B}_{t,1,(1,1)}$.}
    \end{minipage}
    
    % Second row: two subfigures
    \vskip\baselineskip
    \begin{minipage}{0.48\textwidth}
        \centering
        \includegraphics[width=\textwidth]{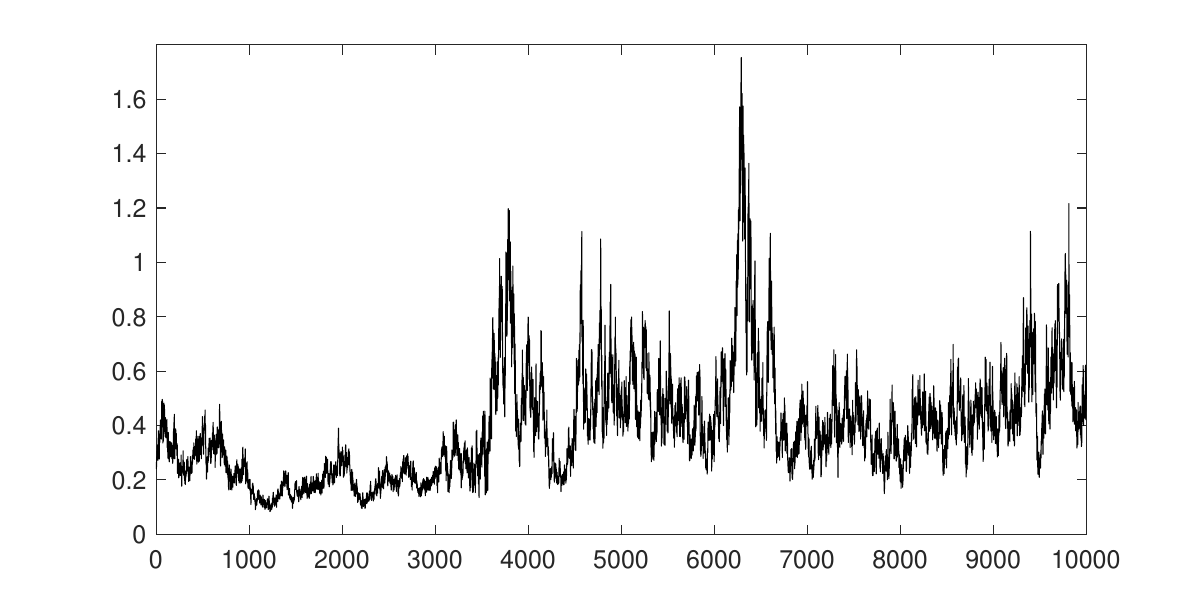} % Add your figure path here
        \subcaption{$\boldsymbol{B}_{2,(1,1)}$.}
    \end{minipage}%
    \hfill
    \begin{minipage}{0.48\textwidth}
        \centering
        \includegraphics[width=\textwidth]{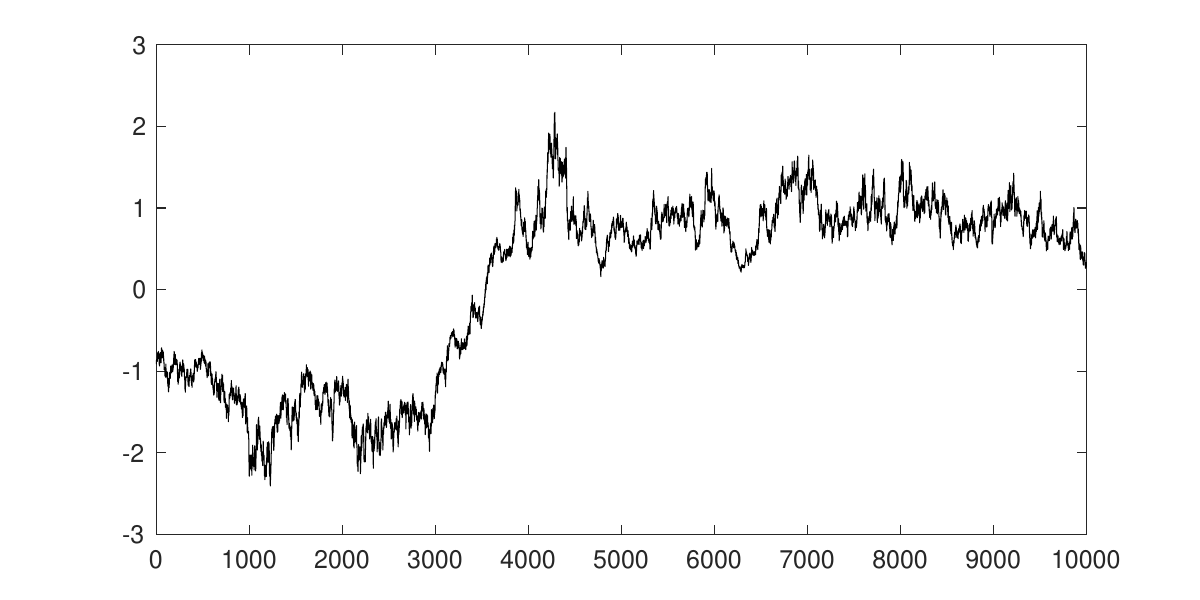} % Add your figure path here
        \subcaption{$\boldsymbol{B}_{3,(1,1)}$.}
    \end{minipage}
    \caption{Trace plots of coefficients and margins sampled using TVP-TVAR(3,1) with a rank of 3.}
    \label{fig:traceplots}
\end{figure}

\begin{figure}
    \centering
    \includegraphics[width=0.6\linewidth]{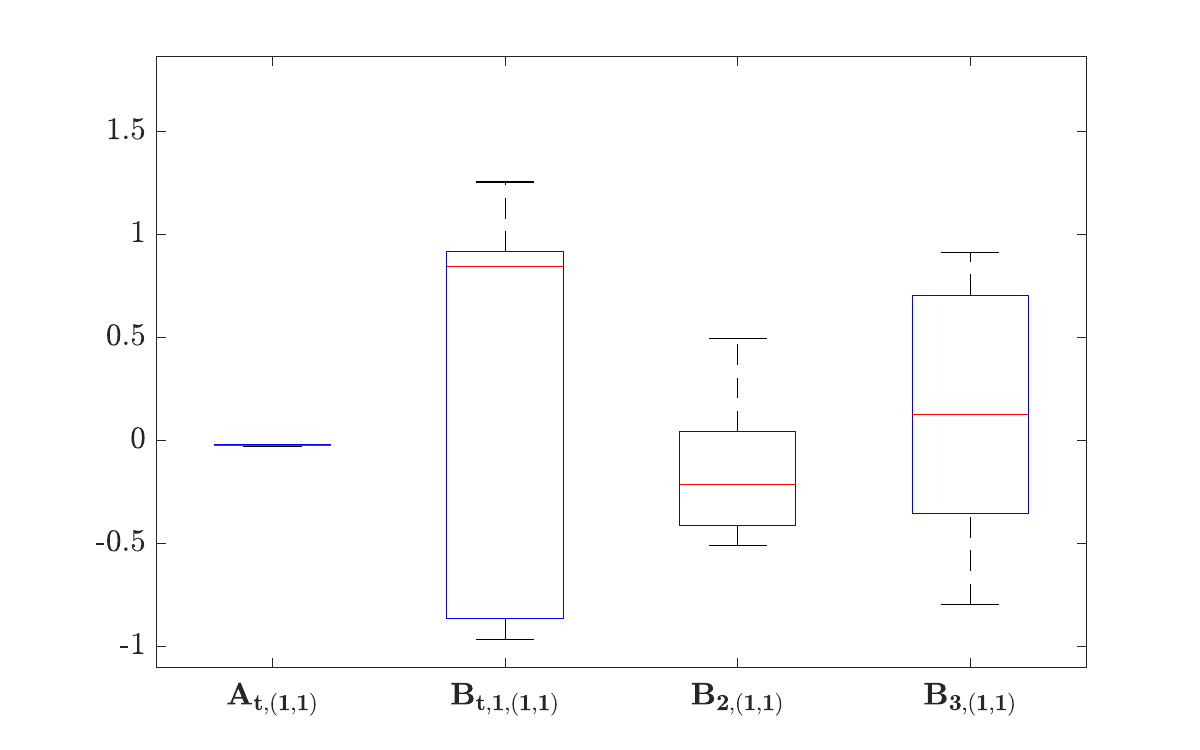}
    \caption{Boxplots of sample mean of coefficients and margins across 10 MCMC runs for the inference of one data set, generated from TVP-TVAR(3,1) with a rank of 3.}
    \label{fig:boxplot}
\end{figure}

Figure \ref{fig:traceplots} and \ref{fig:boxplot} explain why $\text{DIC}^{c,1}$ is more reliable than other DICs. Figure \ref{fig:traceplots} presents the trace plots of the (1,1)-entry in the coefficient matrix $\boldsymbol{A}_{t}$ alongside its decomposed margins - the (1,1) entry in $\boldsymbol{B}_{t,1}$, $\boldsymbol{B}_2$ and $\boldsymbol{B}_3$ - at time point $t=100$. One can also use (1,2) or (1,3) entries in the loadings for demonstration. These trace plots show that the Markov chain of the coefficient mixes well, whereas the margin ones have higher autocorrelation which cannot be solved by thinning. Notably, the sample variance of the coefficient is much lower than those margin counterparts. This autocorrelation, stemming from the indeterminacy of the CP decomposition, also elevates uncertainty of margin sample mean, as presented in Figure \ref{fig:boxplot}. This figure depicts boxplots of sample means of these parameters examined in Figure \ref{fig:traceplots}. Each boxplot displays 10 data points, with each point representing a sample mean derived from one MCMC run. The indeterminacy of the CP decomposition can result in loadings sampled from different MCMC runs representing permuted and sign-switched versions of the same underlying margins. To prevent this phenomenon from distorting our results, we aligned margins across parallel MCMC runs using correlation matching. Specifically, we compute correlations between the (1,1) entry samples shown in Figure \ref{fig:traceplots} and the (1, $r$) entry samples of the same loading in other runs, for $r=1,2,3$. The matching process identifies margins by selecting those with the highest absolute correlation, then adjusts them by multiplying with the sign of the correlation. Figure \ref{fig:boxplot} demonstrates that the boxplot of coefficient sample mean exhibits lower variation compared to the margin counterparts. Since these sample means serve as plug-in parameters ($\hat{\boldsymbol{\theta}}$) in deviance of posterior mean, $\text{DIC}^{c,1}$ yields more reliable results by incorporating coefficient sample means compared to $\text{DIC}^{c,2}$ and $\text{DIC}^{m}$, which use margin sample means.

Next, we illustrate that the DIC can identify true model configurations. Table \ref{tab:model_config} presents the confusion matrix of configuration selection. Applying DICs enables the correct identification of model configurations in nearly all cases. TVAR is only selected when it is the true configuration. When the true configuration involves time-varying parameters, only a handful of data sets are misclassified. While \cite{millar2009comparison} and \cite{chan2016observed} showed that the conditional DIC favors complex models in their simulation studies, we do not reach the same conclusion here. For example, TVAR and TVP-TVAR(3,3) are relatively less complex compared to the other two configurations, but the model selection of the former two attains over 90\% accuracy.

\begin{table}[!htbp]
\footnotesize
\centering
\begin{tabular}{{cc|cccc}}
& & \multicolumn{4}{c}{\small \textbf{Selected}} \\
\multicolumn{2}{c|}{} & TVAR(3) & TVP-TVAR(3,1) & TVP-TVAR(3,2) & TVP-TVAR(3,2) \\
\hline
\multirow[c]{4}{*}{\rotatebox[origin=tr]{90}{\small \textbf{True}}} 
& TVAR(3) & 96 & 0 & 2 & 2 \\
& TVP-TVAR(3,1) & 0 & 97 & 2 & 1 \\
& TVP-TVAR(3,2) & 0 & 8 & 89 & 3 \\
& TVP-TVAR(3,3) & 0 & 3 & 4 & 93 \\

\end{tabular}
\caption{Confusion matrix of configuration selection.}
\label{tab:model_config}
\end{table}
% \begin{figure}[!htbp]
%      \centering
%      \begin{subfigure}[b]{0.3\textwidth}
%          \centering
% \includegraphics[width=\textwidth]{Relative_DIC_TVP_TIV_TIV.pdf}
%          \caption{}
%          \label{fig: }
%      \end{subfigure}
%      \hfill
%      \begin{subfigure}[b]{0.3\textwidth}
%          \centering
%          \includegraphics[width=\textwidth]{Relative_DIC_TIV_TVP_TIV.pdf}
%          \caption{}
%          \label{fig: }
%      \end{subfigure}
%      \hfill
%      \begin{subfigure}[b]{0.3\textwidth}
%          \centering
%          \includegraphics[width=\textwidth]{Relative_DIC_TIV_TIV_TVP.pdf}
%          \caption{}
%          \label{}
%      \end{subfigure}
%         \caption{}
%         \label{fig: }
% \end{figure}
\begin{figure}[!htbp]
     \centering
     \begin{subfigure}[b]{0.35\textwidth}
         \centering
         \includegraphics[width=\textwidth]{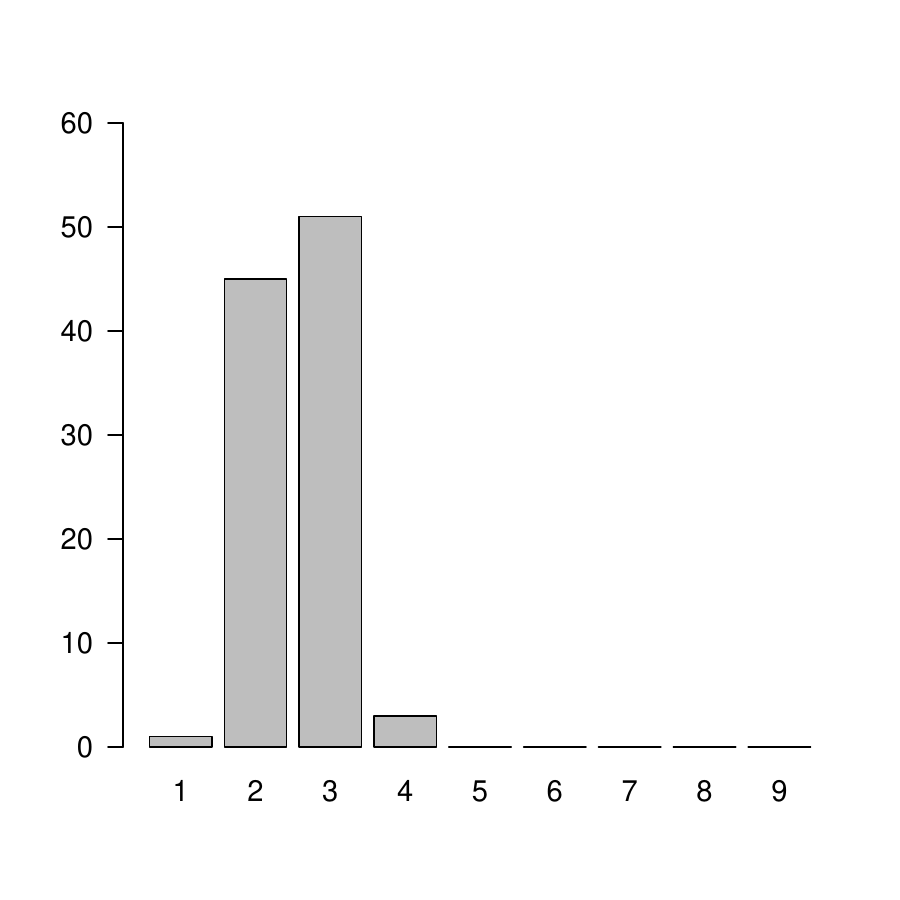}
         \caption{With knee point detection.}
         \label{fig:rank_kneedle}
     \end{subfigure}
     \begin{subfigure}[b]{0.35\textwidth}
         \centering
         \includegraphics[width=\textwidth]{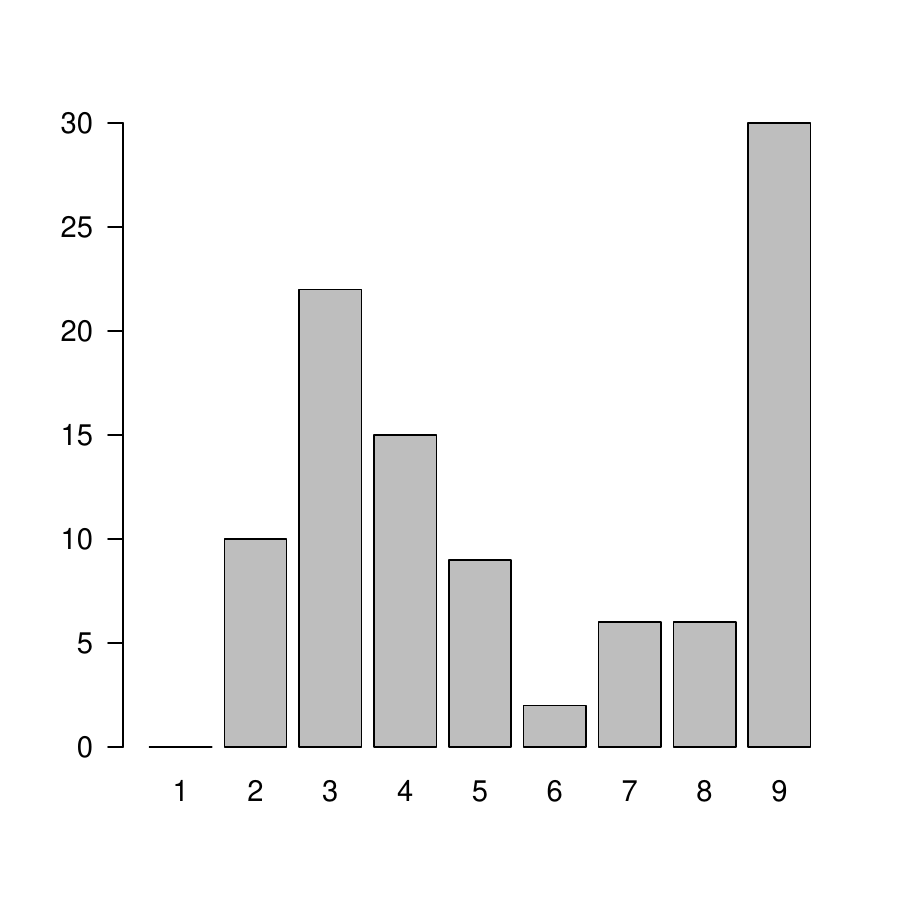}
         \caption{Without knee point detection.}
         \label{fig:rank_min}
     \end{subfigure}
     \hfill
        \caption{Histograms of selected ranks based on the data sets generated from TVP-TVAR(3,1). }
        \label{fig:hist_rank}
\end{figure}

Lastly, we show that the knee point detection improves rank selection. Figure \ref{fig:hist_rank} depicts the distributions of selected ranks if the data is generated from TVP-TVAR(3,1). We record ranks determined using the knee point detection in the left panel and select the ranks corresponding to the lowest DICs in the right panel. According to the left panel, DICs with knee point detection correctly identify the true rank ($R=3$) in over 50 data sets, with almost all remaining data sets selecting ranks adjacent to the true rank (i.e. $R=2$ or $R=4$). In contrast, ranks selected without the knee point detection span from 2 to 9, with $R=9$ having the highest frequency. This suggests that knee point detection alleviates the tendency of DICs to favor overfitted models. Similar conclusions emerge from analysis using other model configurations (see figures in Appendix \ref{appendix:Additional Results in the Monte Carlo Study}).

%%%%%%%%%%%%% real data application %%%%%%%%%%%%%%
%%%%%%%%%%%%%%%%%%%%%%%%%%%%%%%%%%%%%%%
%%%%%%%%%%%%%%%%%%%%%%%%%%%%%%%%%%%%%%%
\section{Empirical Results} \label{Empirical Results}
%%%%%%%%%%%%%%%%%%%%%%%%%%%%%%%%%%%%%%%
\subsection{Data and Implementation}
We apply TVP-TVARs to a functional magnetic resonance
imaging (fMRI) data set about story understanding \citep{wehbe2014simultaneously} to study the time variation of brain connectivity. The data was collected from 8 subjects when reading chapter 9 of \textit{Harry Potter and the Sorcerer’s Stone} \citep{harry}. All subjects are native English speakers and right-handed individuals who are familiar with the story in the chapter. The subject read the chapter in rapid serial visual format, i.e. the token was presented one by one for 0.5 seconds each. The chapter contains approximately 5,200 tokens and was divided into four runs, each lasting approximately 11 minutes. Before and after each run, there would be 20-second and 10-second breaks, respectively, for the subject to stare at a cross on the screen. For each subject in each run, the fMRI data was collected per two seconds from 21764 voxels, corresponding to 117 regions of interest (ROIs).

Following \cite{zhang2021bayesian} and \cite{xiong2023beyond}, we select $N=27$ ROIs which control various cognitive and sensory functions (see the ROIs selected in Appendix \ref{appendix:Data}). For each ROI, we take the average of the voxel data within this particular ROI to form one time series. We discard the data when subjects took breaks and split the time series according to the runs. This process yields 32 (number of subjects $\times$ number of runs) data sets, with each in the format of $T_{\text{run id}}\times 27$ (the average value of $T_{\text{run id}}$ is 323). We then standardize each data set to avoid any scaling issue. 

Following the lag order in \cite{zhang2021bayesian}, we apply TVAR(4), TVP-TVAR(4,$j$), for $j=1,2,3$, to the data sets and select the rank from 1 to 10, extending beyond the rank of 8 previously reported as sufficiently large in \cite{zhang2021bayesian}. For the margin prior, we set $\sigma^2=0.1$, which leads to the variance range of coefficients in $\boldsymbol{A}_p$ in TVAR or $\boldsymbol{A}_{1,p}$ in TVP-TVAR as $\frac{1}{p^2}\text{\texttimes}[10^{-3},10^{-2}]$, for $p=1,\dots,4$, given the rank range. The inverse-Wishart prior of the variance-covariance matrix has parameters $\nu=N+3$ and $\boldsymbol{S}=\mathbf{I}_{N}$. We set $a_k=b_k=0.01$ so that non-zero elements in $\boldsymbol{Q}_j$ have non-informative inverse-gamma priors.
%%%%%%%%%%%%%%%%%%%%%%%%%%%%%%%%%%%%%%%
\subsection{Model Selection} \label{Model Selection}

We first demonstrate the model selection using one data set corresponding to Subject 1 reading the first part of the chapter. Then, we move to a summary of model selection using all 32 data sets. Table \ref{tab:DIC_table} presents the DICs computed with a range of model configurations and ranks. Overall, the DICs corresponding to TVP-TVAR(4,1) are lower than those of other configurations with the same rank value. The rank is selected to be 5 or 6 across configurations, as indicated in bold in the table. By comparing these 4 DICs in bold, we select TVP-TVAR(4,1) as the model configuration with a rank of 6. This suggests that Subject 1, when reading the first part of the chapter, had ROIs that processed information from past signals dynamically and maintained a static framework for gathering information from these signals.

\begin{table}[!htbp]
\footnotesize
\centering
\begin{tabular}{lllllllllll}
\toprule
Model & $R=1$ & $R=2$ & $R=3$ & $R=4$ & $R=5$ & $R=6$ & $R=7$ & $R=8$ & $R=9$ & $R=10$ \\
\hline
TVAR(4)     & 7290.70  & 6660.44 & 6171.06 & 5705.06 & \textbf{5300.59} & 4977.97          & 4746.89 & 4542.20  & 4375.59 & 4243.74 \\
TVP-TVAR(4,1) & 6645.54 & 6096.65 & 5560.65 & 5145.82 & 4762.24          & \textbf{4449.99} & 4309.18 & 4162.92 & 4031.85 & 3909.93  \\
TVP-TVAR(4,2) & 7280.87 & 6637.43 & 6120.45 & 5690.45 & \textbf{5276.53} & 4941.90           & 4697.56 & 4467.36 & 4291.59 & 4115.15 \\
TVP-TVAR(4,3) & 7335.64 & 6781.22 & 6319.24 & 5880.49 & \textbf{5529.93} & 5299.88          & 5024.46 & 4793.37 & 4657.43 & 4590.32 \\
\bottomrule
\end{tabular}
\caption{DICs evaluated from the data for which Subject 1 read the first part of the chapter. DICs corresponding to the knee points are in bold.}
\label{tab:DIC_table}
\end{table}

\begin{figure}[h]
    \centering
    \begin{minipage}{0.48\textwidth}  % Adjust width as needed
            \centering
   \includegraphics[width=\linewidth]{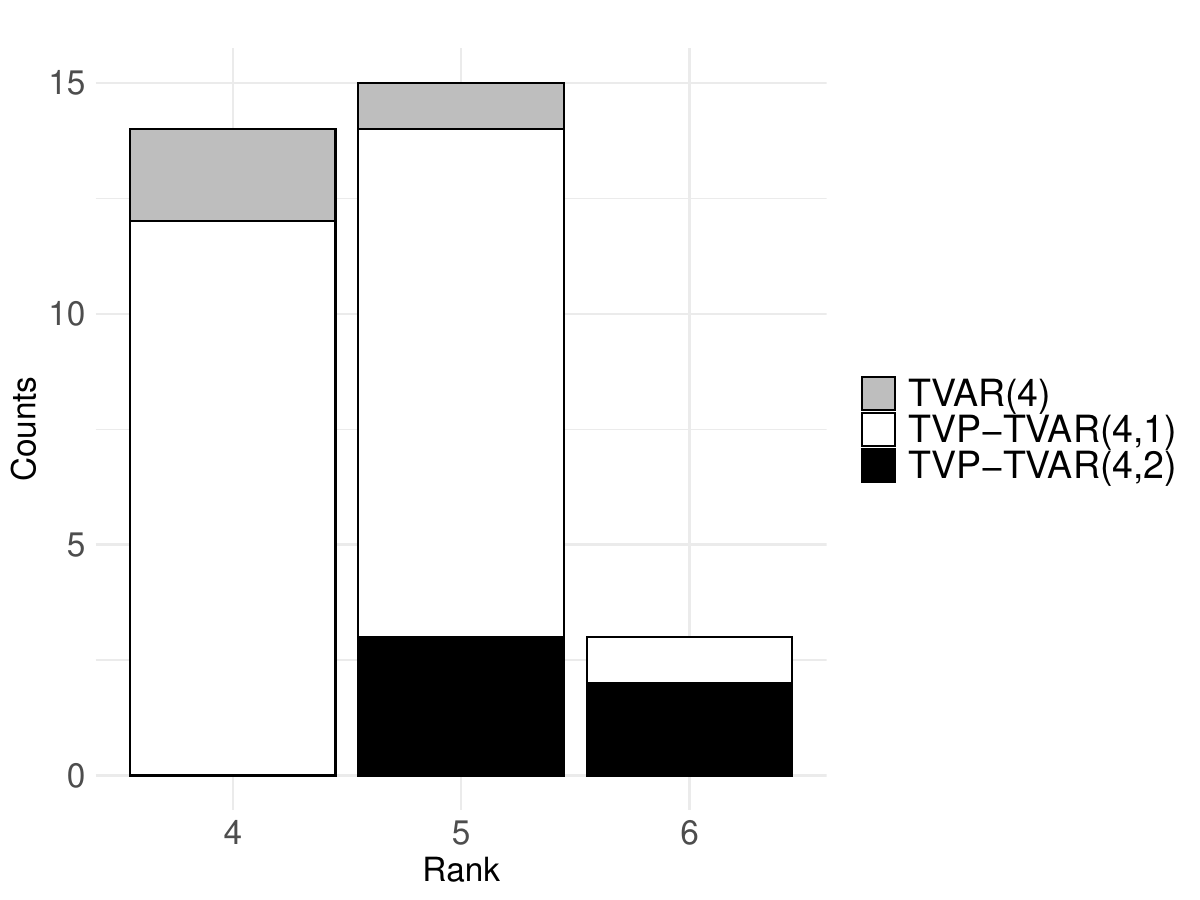}
    \caption{Counts of models selected across 32 data sets. }
    \label{fig:rank selection}
    \end{minipage}
    \hfill
    \begin{minipage}{0.48\textwidth}
        \footnotesize
\centering
\begin{tabular}{{llll}}
\toprule
\input{Model_parameter.tex}
\end{tabular}
\captionof{table}{Averaged number of parameters estimated across all subjects and runs based on different configurations.}
\label{tab:model selection}
    \end{minipage}
\end{figure}

We repeat the same analysis to all 32 data sets, and summarize the results in Figure \ref{fig:rank selection}. Three-quarters of the datasets favor TVP-TVAR(4,1), meaning that VAR coefficients exhibit temporal variation due to evolving signal reception patterns across ROIs. The three TVAR selections correspond to Subject 3 or the first part of the chapter, while TVP-TVAR(4,2) is only selected when the subjects read the second and third parts. None of the data sets identify TVP-TVAR(4,3) as the optimal model fit. Based on these selected ranks, predominantly 4 and 5, Table \ref{tab:model selection} summarizes the average parameter counts for both standard VARs and TVARs. It reveals that both TVAR and TVP-TVARs effectively reduce the number of parameters by over 90\%, inducing parsimonious model structures. Due to the high computational cost in standard VARs, we do not fit the data using these models in the application.

%%%%%%%%%%%%%%%%%%%%%%%%%%%%%%%%%%%%%%%
\subsection{Granger Causality Analysis} \label{Granger Causality Analysis}

Granger causality is a prevalent analysis in neuroscience to identify directional connectivity patterns between brain regions \citep{seth2015granger}. To demonstrate Granger causality analysis using TVP-TVAR, we use the data set of Subject 1 reading the first part. The model implemented is TVP-TVAR(4,1) with a rank of 6, selected according to Table \ref{tab:DIC_table}. The results of other data sets are available upon request.

Informally, Granger causality from time series $m$ to $n$ exists when forecasts of time series $n$ are more accurate by including time series $m$
in the model compared to excluding it. A VAR defines the existence of Granger causality from time series $m$ to $n$ as whether the $(n,m)$ entry in $\boldsymbol{A}_{t,p}$, for $p=1,\dots,P$, are non-zero. If all the associated coefficients are zeros, then time series $m$ does not Granger cause time series $n$. Since Bayesian inference of VARs rarely estimates coefficients as exact zeros without specific priors, such as spike-and-slab prior \citep{mitchell1988bayesian}, we define time series $m$ Granger causes $n$ at time $t$ if $p_{t, (m\rightarrow n)} = p\left(\lvert \boldsymbol{A}_{t,p,(n,m)}\rvert>\delta \text{ for any }p=1,\dots,P\mid \boldsymbol{y}_{1:T}\right)$ is higher than a threshold $p^*$, where we set $\delta$ as 0.01 following \cite{fana2022bayesian} and $p^*$ as 99.9\% to limit the false positive connections. 

\begin{figure}
    \centering
\includegraphics[width=0.35\linewidth]{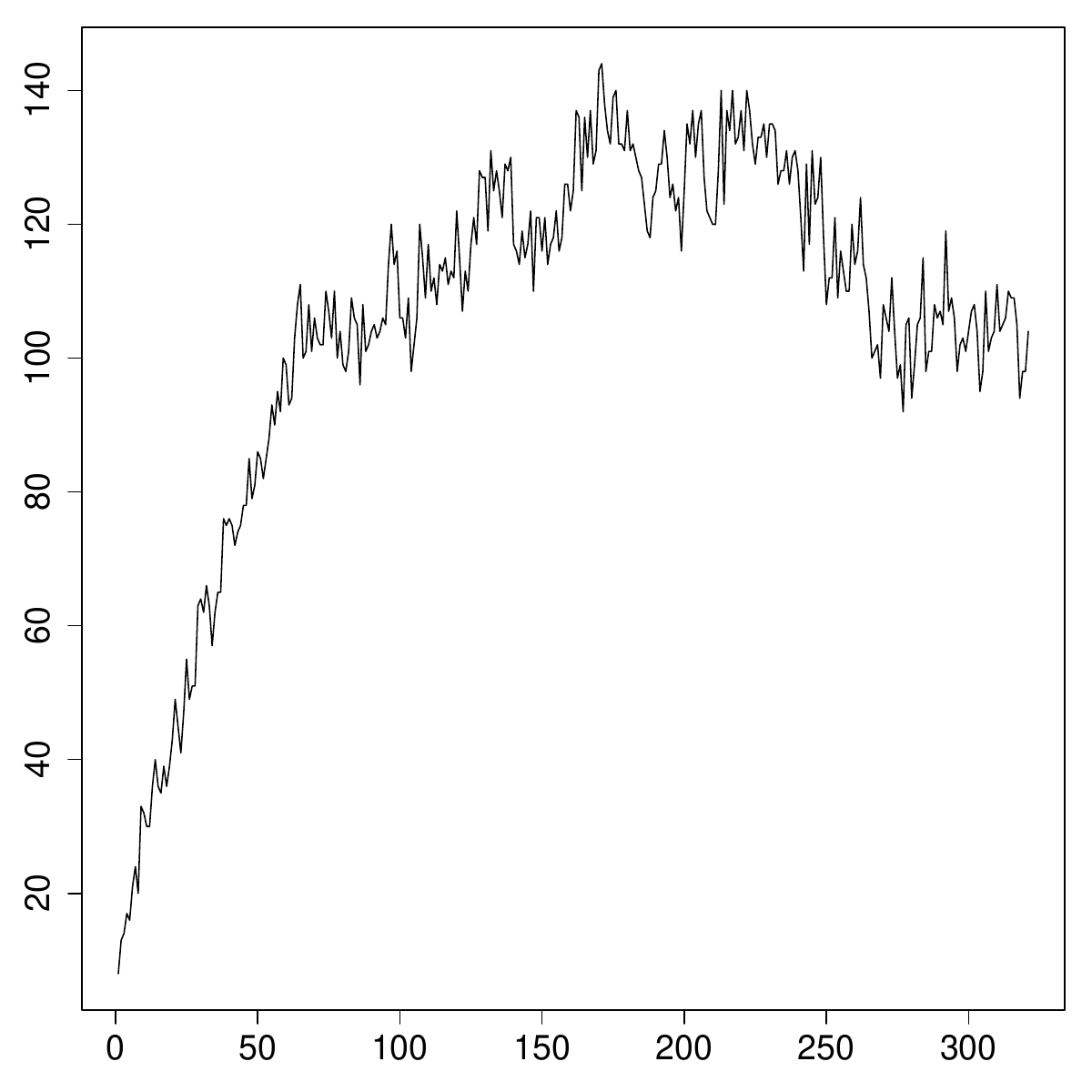}
    \caption{Granger causality time series}
    \label{fig:GC_ts}
\end{figure}

Figure \ref{fig:GC_ts} depicts the number of Granger causalities detected over time. When the subject started reading, there were relatively limited connections between brain regions. Subsequently, the connections increased progressively, maintaining levels above 90 across time points after t=100. This pattern of Granger causality 
is related to the narrative progression in the first part of the chapter. For example, connectivity peaks during a pivotal scene featuring Harry and his friends participating in their first flying lesson. The number of Granger causalities begins to decline around $t=225$ as the narrative focus shifts away from the protagonists.

\begin{figure}[!htbp]
     \centering
     \begin{subfigure}[b]{0.3\textwidth}
         \centering
         \includegraphics[width=\textwidth]{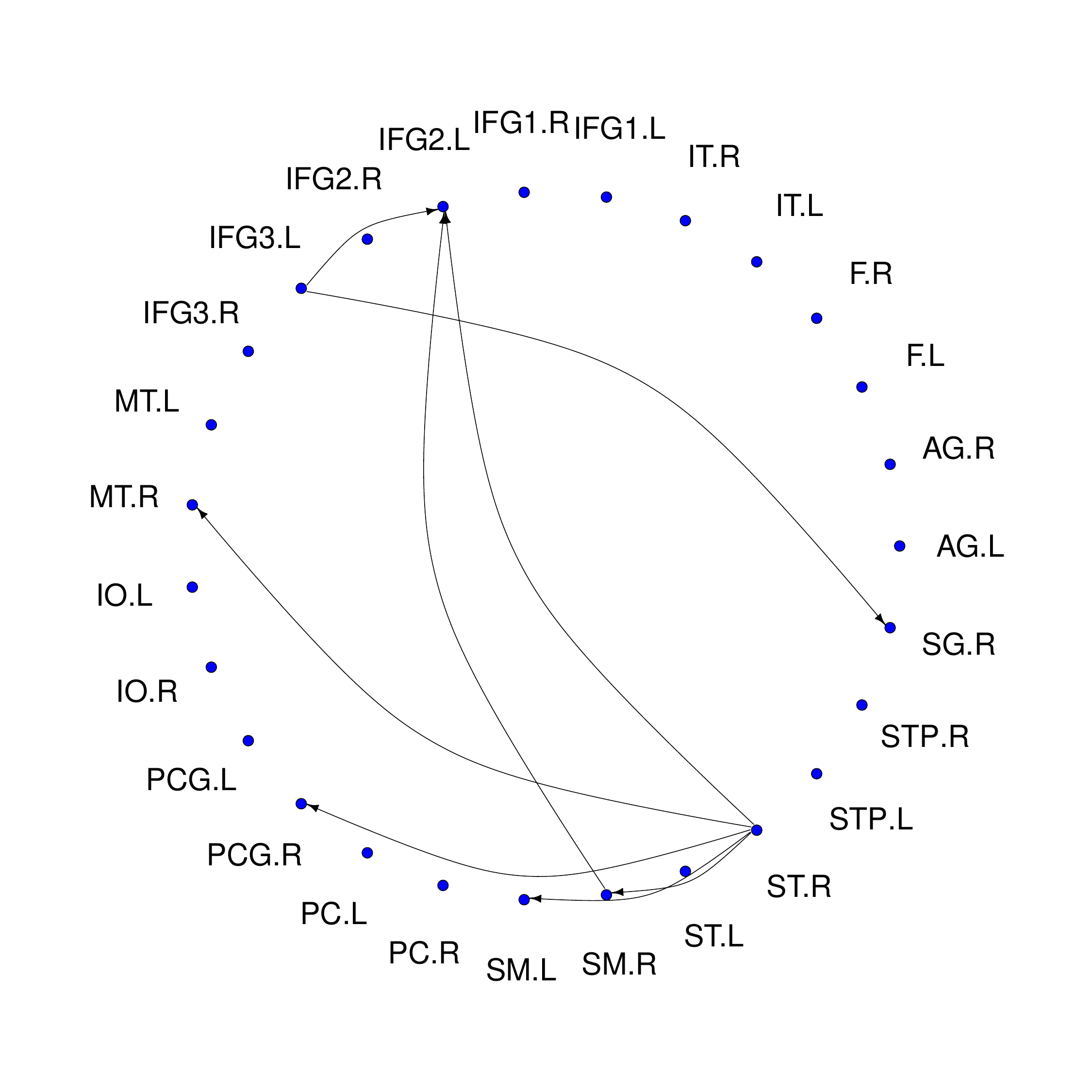}
         \caption{$t=1$}
         \label{fig: GC_1}
     \end{subfigure}
     \hfill
     \begin{subfigure}[b]{0.3\textwidth}
         \centering
         \includegraphics[width=\textwidth]{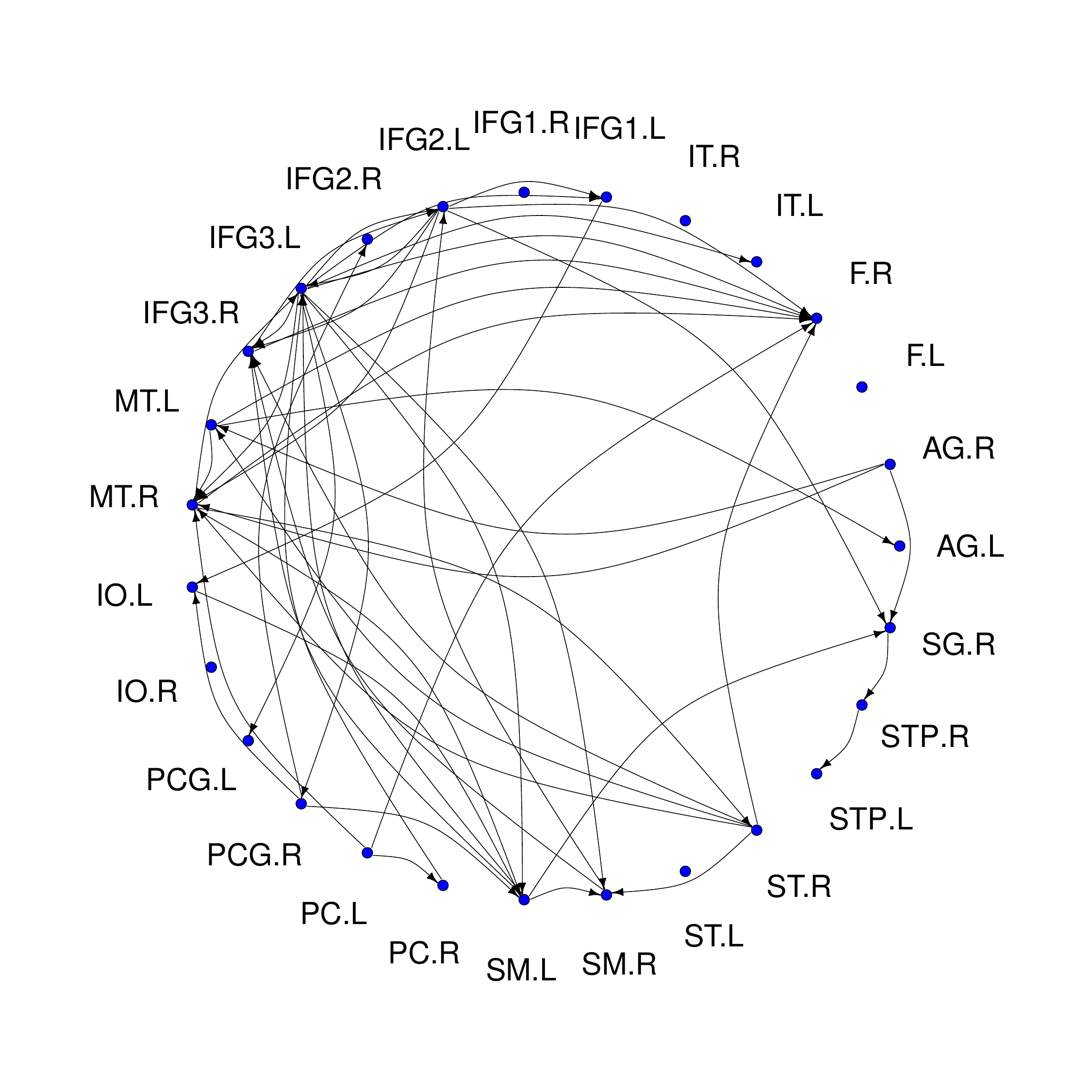}
         \caption{$t=171$.}
         \label{fig: gc_171}
     \end{subfigure}
     \hfill
     \begin{subfigure}[b]{0.3\textwidth}
         \centering
         \includegraphics[width=\textwidth]{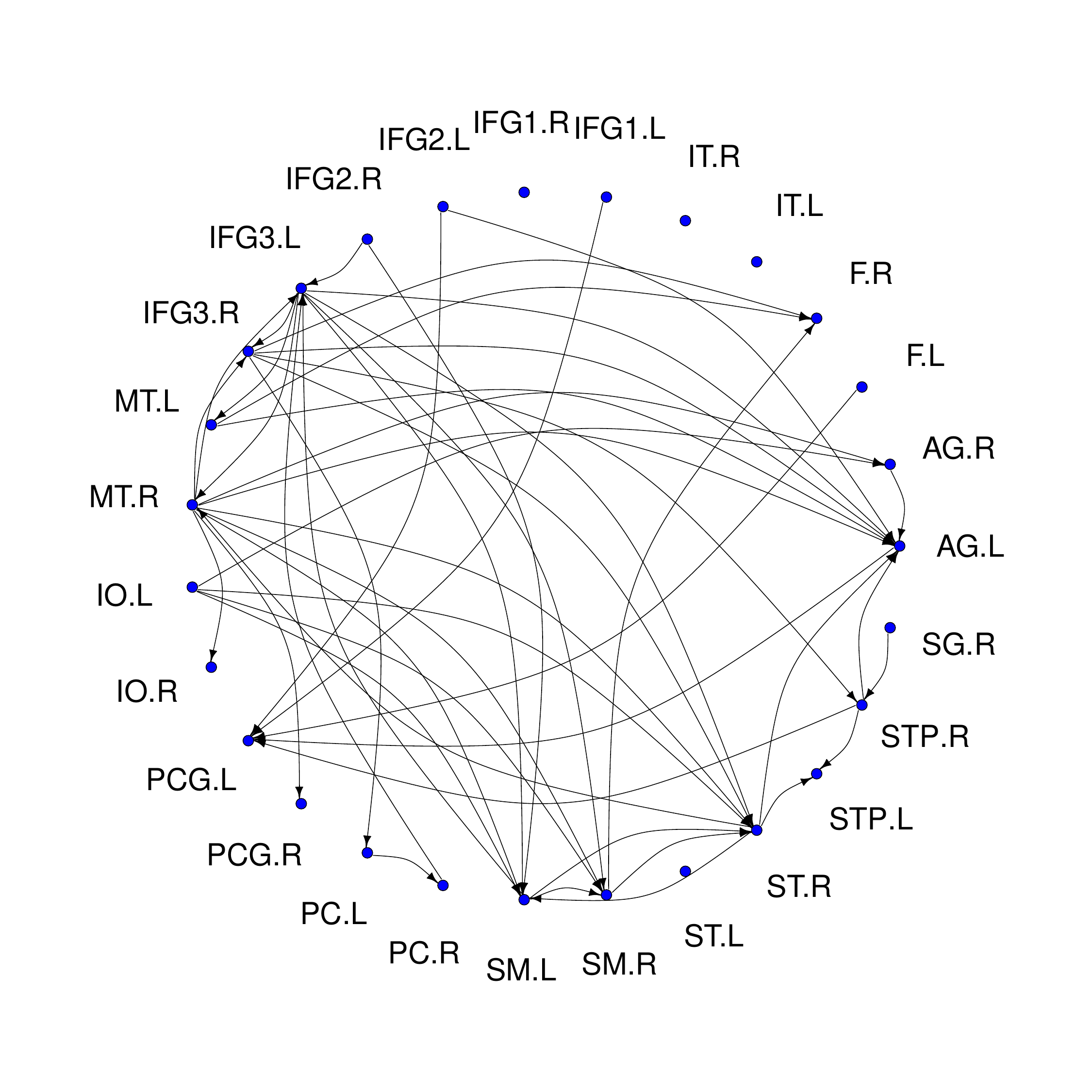}
         \caption{$t=300$}
         \label{fig: gc_300}
     \end{subfigure}
        \caption{Granger causality networks at different time points.}
        \label{fig:GC_networks}
\end{figure}

To illustrate the evolving connectivity patterns of ROIs, we rank $p_{t,(m\rightarrow n)}$ at a particular time point $t$ in descending order and display the first 50 connections for visualization. Figure \ref{fig:GC_networks} provides the Granger causality networks at $t=1$, 171 and 300, corresponding to the start of the chapter, Harry started his flying lesson and the accident of Nievell, respectively. In Figure \ref{fig: GC_1}, ST.R (right superior temporal gyrus) emitted the majority of signals. This region plays an important role in receptive language function and social cognition \citep{bigler2007superior}, indicating that the subject began to process the textual information of this chapter, which involves numerous character interactions. The network became denser in Figure \ref{fig: gc_171} as the story progressed to the flying lesson. The ROI receiving the highest number of signals (7) from other ROIs is the right fusiform gyrus (F.R), which specializes in high-level visual processing functions, including reading and object recognition \citep{weiner2016anatomical}. Compared to Figure \ref{fig: GC_1}, 17 additional connections originated from the inferior frontal gyrus areas (nodes prefixed with IFG), which is a multifunctional region with functionalities including, but not limited to, speech perception, programming sequential order of motor executions and social interaction \citep{liakakis2011diversity}. This pattern aligns with findings reported in \cite{zhang2021bayesian}, who showed similar shifts in Granger causality patterns during another key scene in the chapter. Moving to Figure \ref{fig: gc_300}, the left angular gyrus (AG.L) and the right superior temporal gyrus (ST.R), which are both related to auditory processing \citep{seghier2013angular, bigler2007superior}, not only received more signals than other ROIs but also exhibited the most notable changes compared to the previous figure. In particular, AG.L and ST.R received 9 and 6 signals, respectively, increasing from 1 at $t=171$. A possible explanation for this increase in signals received is that the corresponding story segment involves more auditory elements, such as teacher instructions and laughter, compared to other story segments.
% \begin{figure}
%     \centering
%     \begin{minipage}{0.3\textwidth}
%         \centering
%         \includegraphics[width=\textwidth]{GC_1.pdf}
%         \caption{Caption 1}
%     \end{minipage}
%     \begin{minipage}{0.3\textwidth}
%         \centering
%         \includegraphics[width=\textwidth]{GC_171.pdf}
%         \caption{Caption 2}
%     \end{minipage}
%     \begin{minipage}{0.3\textwidth}
%         \centering
%         \includegraphics[width=\textwidth]{GC_300.pdf}
%         \caption{Caption 3}
%     \end{minipage}
%     \caption{Granger causality networks}
% \end{figure}

%%%%%%%%%%%%% conclusion %%%%%%%%%%%%%%
%%%%%%%%%%%%%%%%%%%%%%%%%%%%%%%%%%%%%%%
%%%%%%%%%%%%%%%%%%%%%%%%%%%%%%%%%%%%%%%
\vspace{-0.2cm}
\section{Conclusion and Discussion} \label{sec7}
We propose the time-varying parameter tensor vector autoregression with three model configurations and implement a conditional DIC with knee point detection for model selection. The fMRI data application demonstrates that time variation in the response loading is the preferred configuration for most subjects. This finding supports the non-stationarity of fMRI data and reveals time-varying dynamics between the data and its representation of past information.

Our work can be extended in several directions. First, it would be interesting to incorporate TVP-TVARs with shrinkage priors to allow a hybrid model analogous to the VAR counterpart proposed in \cite{chan2023large}. One caveat is the identifiability issue of the margins, which could hinder sampling and interpretation. Second, developing techniques for obtaining convergent margin Markov chains can potentially lower the uncertainty in the alternative DICs specified in Section \ref{Model and Rank Selection}. Third, heteroskedasticity plays a critical role in VAR applications within econometrics \citep{clark2023stochastic}, suggesting that adding heteroskedasticity such as stochastic volatility to TVP-TVARs is worth investigating. 
%%%%%%%%%%%%% ackonwledgments %%%%%%%%%%%%%%
%\section*{Acknowledgments}

%%%%%%%%%%%%%%%%%%%%%%%%%%%%%%%%%%%%%%%%%%%

% \bibliographystyle{apalike}  
% \bibliography{references} 
%%%%%%%%%%%%%%%%%%%%%%%%%%%%%%%%%%%%%%%%%%
%%%%%%%%%%%%%%%%%%%%%%%%%%%%%%%%%%%%%%%%%%
\newpage

\appendix
\begin{center}
{\Large SUPPLEMENTARY MATERIAL}
\end{center}
\section{Basic Notations and Definitions}
\label{appendix:Basic Notations and Definitions}
A tensor $\boldsymbol{\mathcal{A}}$ is a $J$-th-order tensor if $\boldsymbol{\mathcal{A}}\in\mathbb{R}^{I_1\times\cdots\times I_J}$ with entries $\boldsymbol{\mathcal{A}}_{i_1,\dots,i_J}$ for $i_j=1,\dots,I_j$. Vectors and matrices correspond to first- and second-order tensors, respectively.
Similar to the definition of a diagonal matrix, a tensor is called a $J$-th-order \textit{superdiagonal} tensor if $I_1=\cdots =I_J=I$ and only its ($i,\, \dots,\,i$) entries are non-zero for $i=1,\,\dots,\,I$. Extracting entries by a selected index from a matrix $\boldsymbol{A}$ and a tensor $\boldsymbol{\mathcal{A}}$ is akin. 

Matricization of a $J$-th-order tensor ($J$\textgreater 2) is an operation that transforms the tensor into a matrix. 
There are $J$ possible matricizations for a $J$-th-order tensor $\boldsymbol{\mathcal{A}}$, and the matricization to the $j$-th dimension is called the \textit{mode-j matricization} with notation $\boldsymbol{\mathcal{A}}_{(j)}\in\mathbb{R}^{I_j\times \prod_{j^\prime\neq j}^J I_{j^\prime}}$, where the $i_j$-th row corresponds to the vectorization of $\boldsymbol{\mathcal{A}}_{\dots,i_j,\dots}$.

The outer (or tensor) product of two vectors $\boldsymbol\beta_1\in\mathbb{R}^{I_1}$ and $\boldsymbol\beta_2\in\mathbb{R}^{I_2}$ is $\boldsymbol\beta_1\circ\boldsymbol\beta_2$, yielding an $I_1$-by-$I_2$ matrix with its ($i_1$, $i_2$) entry as $\boldsymbol\beta_{1,i_1}\boldsymbol\beta_{2,i_2}$. Assume we have $J$ vectors with $\boldsymbol\beta_j\in\mathbb{R}^{I_j}$ denoting the $j$-th vector, then the \textit{$J$-way outer product} of these vectors, $\boldsymbol\beta_1\circ\cdots\circ\boldsymbol\beta_J$, is a $I_1\times\cdots\times I_J$ tensor, with ($i_1,\, \dots ,\, i_j$) entry as $\boldsymbol\beta_{1,i_1}\cdots\boldsymbol\beta_{J,i_J}$. Note that the outer product can also apply to a pair of a vector and a matrix. Assume that the vector is $\boldsymbol{a}\in\mathbb{R}^{K}$ and the matrix is $\boldsymbol{B}\in\mathbb{R}^{M\times N}$. The their outer product, $\boldsymbol{a}\circ\boldsymbol{B}$, is a third-order tensor with dimension $K$\texttimes$M$\texttimes$N$, and the $(k, m, n)$ entry equals to $\boldsymbol{a}_k\boldsymbol{B}_{m.n}$.

\section{Marginal DIC in TVP-TVAR}
\label{appendix:Marginal DIC}
The key term in the marginal DIC is $p\left(\boldsymbol{y}_{1:T}\mid \boldsymbol{b}_{j^\prime},\boldsymbol{\Omega}, \boldsymbol{Q}_j\right)$, which marginalizes $\boldsymbol{b}_{1:T,j}$:
\begin{equation}
    \int p\left(\boldsymbol{y}_{1:T}\mid \boldsymbol{b}_{1:T,j}, \boldsymbol{b}_{j^\prime},\boldsymbol{\Omega}, \right)p\left(\boldsymbol{b}_{1:T,j}\mid \boldsymbol{Q}_j\right)d \boldsymbol{b}_{1:T,j}.
    \label{marginal_DIC}
\end{equation}
Analogous to \cite{chan2018bayesian}, $p\left(\boldsymbol{y}_{1:T}\mid \boldsymbol{b}_{j^\prime},\boldsymbol{\Omega}, \boldsymbol{Q}_j\right)$ has a closed form because of the state-space representations described in Section \ref{Posterior Computation}. For $j=1$ or 3, $\boldsymbol{y}\mid \boldsymbol{b}_{1:T,j}, \boldsymbol{b}_{j^\prime},\boldsymbol{\Omega}\sim\mathcal{N}\left(\boldsymbol{Z}_j\boldsymbol{b}_j, \mathbf{I}_{T}\otimes \boldsymbol{\Omega}\right)$, where $\boldsymbol{y}=\left(\boldsymbol{y}^\prime_1,\dots, \boldsymbol{y}^\prime_T\right)^\prime$, $\boldsymbol{b}_j=\left(\boldsymbol{b}^\prime_{1,j},\dots, \boldsymbol{b}^\prime_{T,j}\right)^\prime$, $\boldsymbol{Z}_j=\text{diag}\left(\boldsymbol{Z}_{1,j},\dots,\boldsymbol{Z}_{T,j}\right)$; $\boldsymbol{b}_j\mid \boldsymbol{Q}_j\sim\mathcal{N}\left(\mathbf{0},\boldsymbol{H}^{-1}\boldsymbol{S}_j\left(\boldsymbol{H}^{-1}\right)^\prime\right)$, where {\setlength{\arraycolsep}{2pt}
\renewcommand{\arraystretch}{0.5}$\boldsymbol{H}=\begin{pmatrix} 
    \mathbf{I}_{I_jR} & \mathbf{0} & \dots & \mathbf{0} \\
    -\mathbf{I}_{I_jR} & \mathbf{I}_{I_jR} & \ddots & \mathbf{0}\\
    \vdots &   \ddots & \ddots    & \vdots\\
    \mathbf{0} & \dots & -\mathbf{I}_{I_jR} & \mathbf{I}_{I_jR}
    \end{pmatrix}$}, $\boldsymbol{S}_j=\text{diag}\left(\boldsymbol{\Sigma}_j,\boldsymbol{Q}_j,\dots,\boldsymbol{Q}_j \right)$. By plugging in the probability density functions of the above two distributions to \labelcref{marginal_DIC}, we can get
\renewcommand{\arraystretch}{1}
\begin{align*}
    p\left(\boldsymbol{y}_{1:T}\mid \boldsymbol{\theta}_{\textbackslash \boldsymbol{b}_{1:T,j}}\right) & = (2\pi)^{-\frac{TN}{2}}\lvert \boldsymbol{\Omega}\rvert^{-\frac{T}{2}}\lvert \boldsymbol{Q}_j\rvert^{-\frac{T-1}{2}}\lvert \boldsymbol{\Sigma}_j\rvert^{-\frac{1}{2}}\lvert \boldsymbol{V}_j\rvert^{-\frac{1}{2}}\exp\Bigl\{-\frac{1}{2}\left[\boldsymbol{y}^\prime\left(\mathbf{I}_T\otimes \boldsymbol{\Omega}\right)^{-1}\boldsymbol{y}-\mathbf{m}^\prime_j\boldsymbol{V}^{-1}_j\boldsymbol{m}_j\right]\Bigl\},
\end{align*}
where $\boldsymbol{V}_j=\boldsymbol{Z}^\prime_j\left(\mathbf{I}_T\otimes \boldsymbol{\Omega}\right)^{-1}\boldsymbol{Z}_j+\boldsymbol{H}^\prime\boldsymbol{S}^{-1}_j\boldsymbol{H}$, $\boldsymbol{m}_j=\boldsymbol{Z}^\prime_j\left(\mathbf{I}_T\otimes \boldsymbol{\Omega}\right)^{-1}\boldsymbol{y}$. Thus, $\log p\left(\boldsymbol{y}_{1:T}\mid \boldsymbol{b}_{j^\prime},\boldsymbol{\Omega},\boldsymbol{Q}_j\right)$ is written as
\begin{align*}
    &-\frac{TN}{2}\log 2\pi -\frac{T}{2}\log \lvert \boldsymbol{\Omega}\rvert - \frac{T-1}{2}\log \lvert \boldsymbol{Q}_j\rvert- \frac{1}{2}\log \lvert \boldsymbol{\Sigma}_j\rvert-\frac{1}{2}\log \lvert \boldsymbol{V}_j\rvert-\frac{1}{2}\left[\boldsymbol{y}^\prime\left(\mathbf{I}_T\otimes \boldsymbol{\Omega}\right)^{-1}\boldsymbol{y}-\mathbf{m}^\prime_j\boldsymbol{V}^{-1}_j\boldsymbol{m}_j\right].
\end{align*}
For $j=2$, the expression of $\log p\left(\boldsymbol{y}_{1:T}\mid \boldsymbol{b}_{j^\prime},\boldsymbol{\Omega},\boldsymbol{Q}_j\right)$ is the same after modifying $\boldsymbol{Q}_2$ according to the order of elements in $\boldsymbol{b}^*_{t,2}$. We apply the algorithm in Section \ref{Posterior Computation} to sample unknown parameters to approximate the posterior mean deviance. 

\section{Additional Results in the Monte Carlo Study}
\label{appendix:Additional Results in the Monte Carlo Study}

\begin{figure}[!htbp]
     \centering
     \begin{subfigure}[b]{0.33\textwidth}
         \centering
         \includegraphics[width=\textwidth]{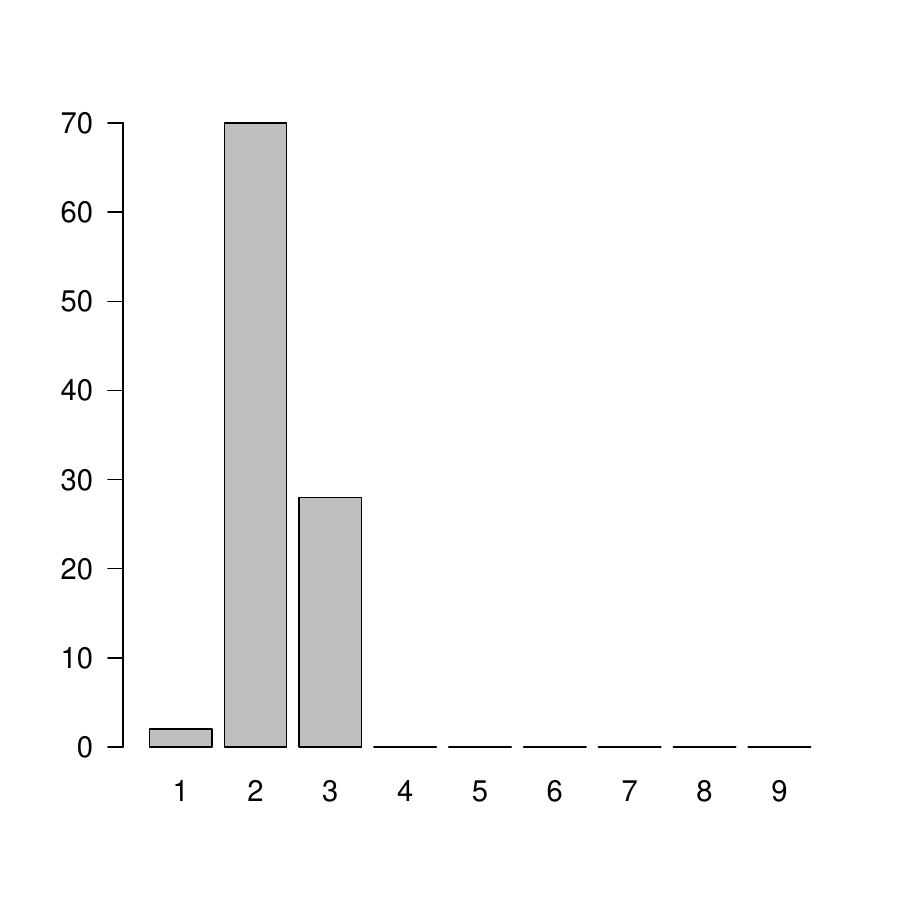}
         \caption{With knee point detection.}
         \label{fig:rank_kneedle}
     \end{subfigure}
     \begin{subfigure}[b]{0.33\textwidth}
         \centering
         \includegraphics[width=\textwidth]{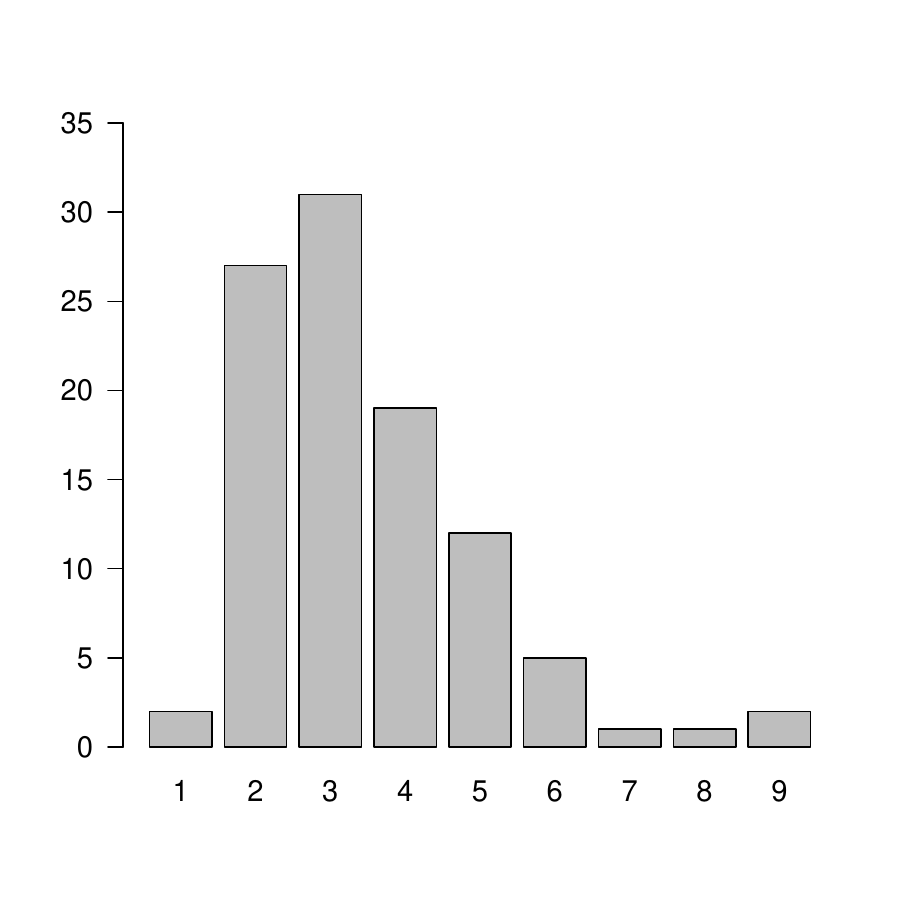}
         \caption{Without knee point detection.}
         \label{fig:rank_min}
     \end{subfigure}
     \hfill
        \caption{Histograms of selected ranks based on data sets generated from TVAR(3). }
        \label{fig:hist_rank_TIV_TVAR}
\end{figure}

\begin{figure}[!htbp]
     \centering
     \begin{subfigure}[b]{0.33\textwidth}
         \centering
         \includegraphics[width=\textwidth]{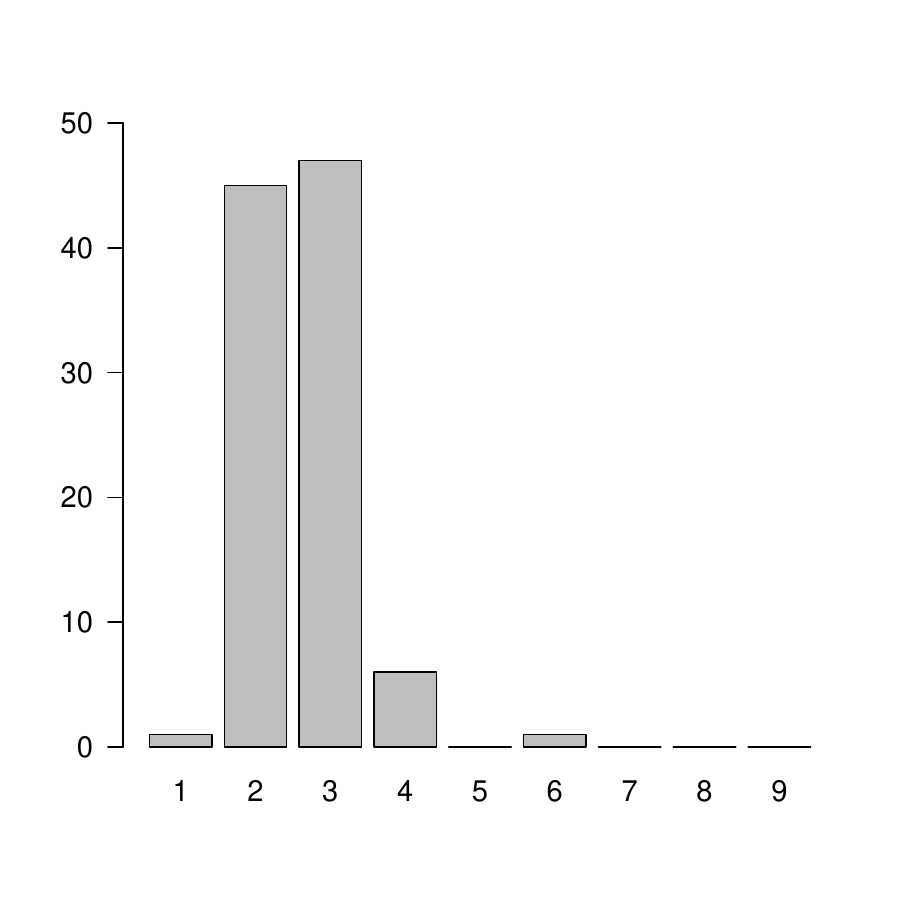}
         \caption{With knee point detection.}
         \label{fig:rank_kneedle}
     \end{subfigure}
     \begin{subfigure}[b]{0.33\textwidth}
         \centering
         \includegraphics[width=\textwidth]{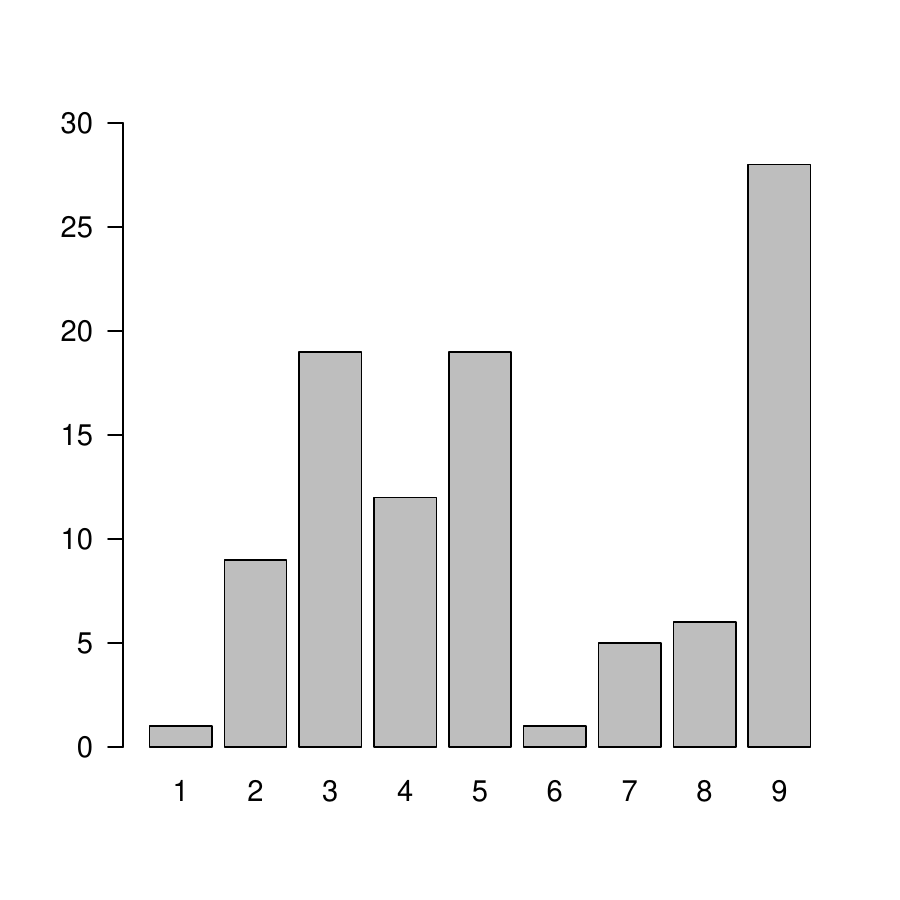}
         \caption{Without knee point detection.}
         \label{fig:rank_min}
     \end{subfigure}
     \hfill
        \caption{Histograms of selected ranks based on data sets generated from TVP-TVAR(3,2). }
        \label{fig:hist_rank_TIV_TVAR}
\end{figure}

\begin{figure}[!htbp]
     \centering
     \begin{subfigure}[b]{0.33\textwidth}
         \centering
         \includegraphics[width=\textwidth]{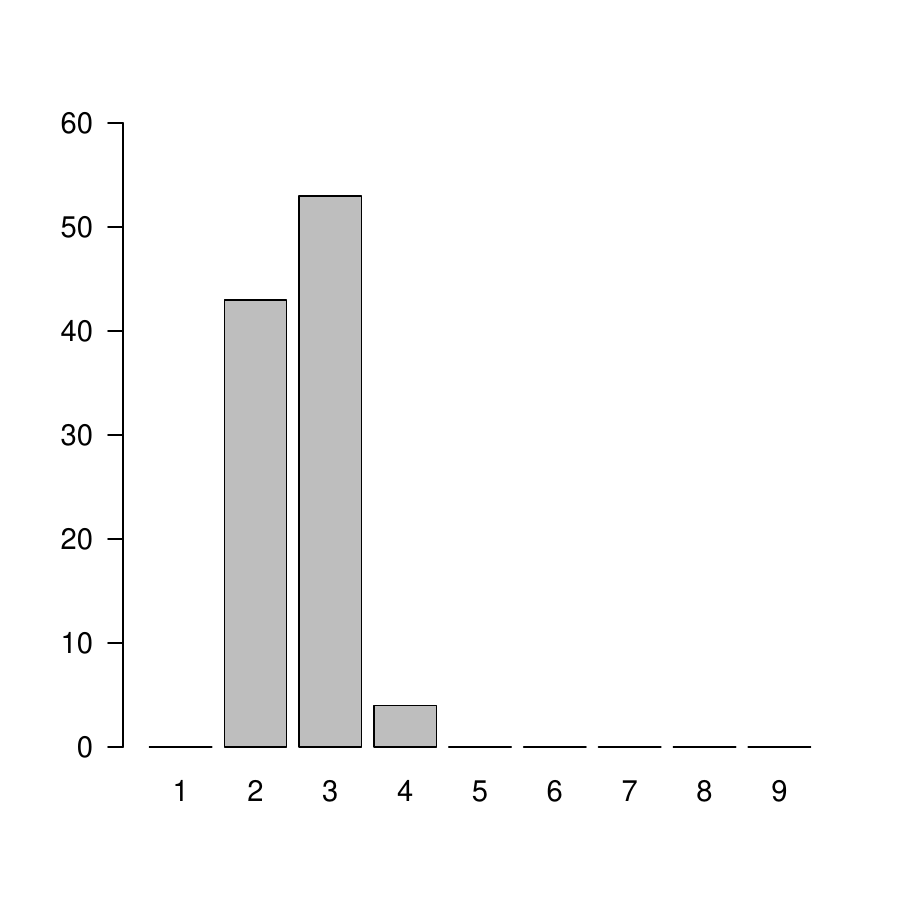}
         \caption{With knee point detection.}
         \label{fig:rank_kneedle}
     \end{subfigure}
     \begin{subfigure}[b]{0.33\textwidth}
         \centering
         \includegraphics[width=\textwidth]{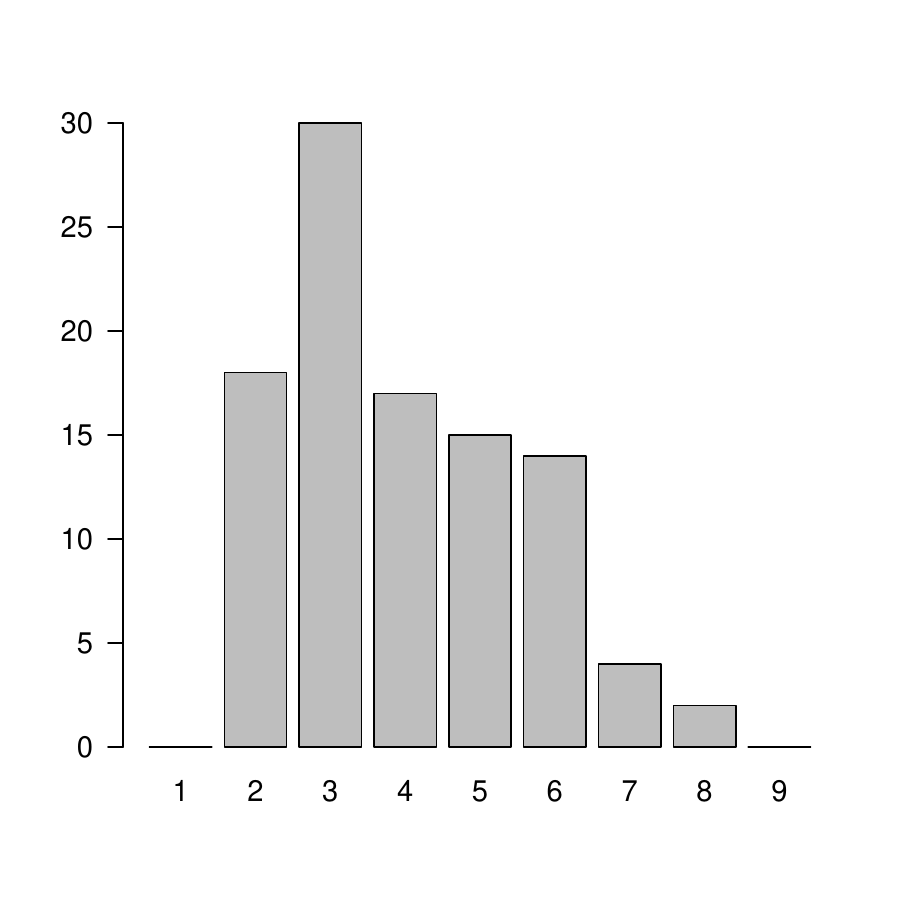}
         \caption{Without knee point detection.}
         \label{fig:rank_min}
     \end{subfigure}
     \hfill
        \caption{Histograms of selected ranks based on data sets generated from TVP-TVAR(3,3). }
        \label{fig:hist_rank_TIV_TVAR}
\end{figure}

\section{Data}
\label{appendix:Data}
\vspace{-1cm}
\begin{center}
\begingroup
\small
\begin{longtable}{cll}%
\toprule
& ROI                                     & Label \\
\hline
1                    & Angular gyrus                           & AG    \\
2                    & Fusiform gyrus                          & F     \\
3                    & Inferior temporal gyrus                 & IT    \\
4                    & Inferior frontal gyrus, opercular part  & IFG 1 \\
5                    & Inferior frontal gyrus, orbital part    & IFG 2 \\
6                    & Inferior frontal gyrus, triangular part & IFG 3 \\
7                    & Middle temporal gyrus                   & MT    \\
8                    & Inferior occipital gyrus                & IO    \\
9                    & Precental gyrus                         & PCG   \\
10                   & Precuneus                               & PC    \\
11                   & Supplementary motor area                & SM    \\
12                   & Superior temporal gyrus                 & ST    \\
13                   & Superior Temporal pole                  & STP   \\
14                   & Supramarginal gyrus                     & SG   \\
\bottomrule
\caption{27 regions of interest from both right and left  cerebral hemispheres, except the supramarginal gyrus for which only the right hemisphere was considered.}
\label{data appendix}
\end{longtable}
\endgroup

\end{center}

\end{document}